\documentclass[prX,twocolumn,amsmath,amssymb]{revtex4}
\usepackage{graphicx}
\usepackage{latexsym}
\usepackage{txfonts}
\usepackage{amsmath,amssymb} 
\usepackage{graphicx}
\usepackage{amsmath}
\usepackage{amssymb}
\usepackage{amsfonts}
\usepackage{upgreek}
\usepackage{txfonts}
\usepackage{color}
\usepackage{comment}
\usepackage{rotating}
\usepackage{mathdots}
\begin{document}

\title{
Fluctuations of imbalanced fermionic superfluids in two dimensions induce\\
continuous quantum phase transitions and non-Fermi liquid behavior
}

\author{Philipp Strack}
\affiliation{Department of Physics, Harvard University, Cambridge MA 02138, USA}

\author{Pawel Jakubczyk}
\affiliation{Institute of Theoretical Physics, Faculty of Physics, University of Warsaw, Ho\.za 69, 00-681 Warsaw, Poland}

\date{\today}

\begin{abstract}
We study the nature of superfluid pairing in imbalanced Fermi mixtures in two 
spatial dimensions. We present evidence that the combined effect of 
Fermi surface mismatch and order parameter fluctuations of the superfluid condensate 
can lead to continuous quantum phase transitions from a normal Fermi mixture to an intermediate 
Sarma-Liu-Wilczek superfluid with two gapless Fermi surfaces -- even when mean-field theory (incorrectly) 
predicts a first order transition to a phase-separated "Bardeen-Cooper-Schrieffer plus excess fermions" 
ground state.
We propose a mechanism for non-Fermi liquid behavior from 
repeated scattering processes between the two Fermi surfaces and 
fluctuating Cooper pairs. Prospects for experimental observation with 
ultracold atoms are discussed.

\end{abstract}

\maketitle

\section{Introduction}
Understanding the nature of strongly correlated fermion 
systems, that do not obey the rules of Landau's Fermi liquid \cite{landau56,nozieres64}, 
remains  one of the most pressing issues in the physics of correlated matter. 
The unifying property of non-Fermi liquid states are the short lifetimes 
of quasi-particles. This makes 
the physics interesting and the theoretical description hard. 
In strong non-Fermi liquids, there are no well-defined quasiparticles at all, that is, there are no electronic excitations whose 
lifetime becomes 
a large timescale when asymptoting energies toward zero. 
In weak non-Fermi liquids, the interactions of fermions with fluctuations 
leads to reduced lifetimes and anomalous thermodynamics but ultimately well-defined quasiparticles. 

Non-Fermi liquids can occur in a variety of physical contexts: ranging 
from ceramic copper oxides and transition metal compounds at below-room-temperatures 
to quantum chromodynamics at finite quark density in conditions similar to those right 
after the big bang (temperatures of millions of Kelvins) \cite{schofield99,varma02,schaefer04,loehneysen07}.
A popular and experimentally relevant scenario for non-Fermi liquid behavior is the proximity 
to a quantum critical point or phase \cite{schofield05,sachdev11}. The crossover from Fermi liquid to 
non-Fermi liquid behavior in solid state materials can be induced for example, 
by changing chemical composition, pressure, or straining the crystal lattice.

Until not so long ago, the best known examples of Fermi liquids accessible 
in table-top experiments were $^3$He \cite{greywall83,vollhardt03} 
and normal metals with low ordering temperatures.
Experiments with ultracold atomic quantum gases \cite{zwerger08} provide a 
complementary route to study interacting Fermi systems, sometimes in novel regimes 
beyond what is possible in solid state situations. Using two-body 
Feshbach resonances, it is now possible to 
prepare interacting Fermi gases of (electrically) charge-neutral atoms with tunable interactions 
in odd-integer, nuclear hyperfine states in the quantum degeneracy regime, 
that is, at nano-Kelvin temperatures corresponding to fractions of the 
Fermi temperature \cite{jin99,zwierlein03,koehl05}.

Density- and/or mass-imbalanced Fermi mixtures 
\cite{zwierlein06,partridge06,grimm12} represent such a new beyond-solid-state synthetic quantum material.
A bold proposition by Liu and Wilczek \cite{liu03} on pairing in such fermion 
mixtures invoked a homogeneous superfluid coexistent with 
``metallic" fermions on two Fermi surfaces
triggered research on Sarma's earlier proposal \cite{sarma63} as well as various generalization 
into the modern context of ultracold Fermi gases
\cite{yip03,bedaque03,liu04,schwenk06,parish07,he08,pilati08,fisher09,randeria10,sheehy10, cichy11}.

The common point of departure for these analysis is the 
attractively interacting (spin-) polarized Fermi gas
\begin{align}
H=&\sum_{\mathbf{k}\sigma}
\xi_{\mathbf{k}\sigma}c^\dagger_{\mathbf{k}\sigma}
c_{\mathbf{k}\sigma}
+
\frac{g}{V}\sum_{\mathbf{k},\mathbf{k}',\mathbf{q}} 
c^{\dagger}_{\mathbf{k}+\mathbf{q}/2,\uparrow} 
c^{\dagger}_{-\mathbf{k}+\mathbf{q}/2,\downarrow} 
c_{\mathbf{k}'+\mathbf{q}/2,\downarrow} 
c_{-\mathbf{k}'+\mathbf{q}/2,\uparrow} \;,
\label{eq:hamiltonian}
\end{align}
where the dispersion of the 
two spin-components is $\xi_{\mathbf{k}\sigma}=\frac{\mathbf{k}^2}{2m_{\sigma}}-\mu_\sigma$ with 
$\sigma=\uparrow$, $\downarrow$, 
and $g<0$ is an attractive, structureless contact interaction between the species. Each dispersion is inflection 
symmetric $\xi_{\mathbf{k}\sigma}=\xi_{-\mathbf{k}\sigma}$. $V$ is the volume and 
in general the masses $m_{\sigma}$ and chemical potentials $\mu_{\sigma}$ of each spin species 
can be different.

\begin{figure}[t]
\vspace*{1mm}
\includegraphics*[width=42mm,angle=0]{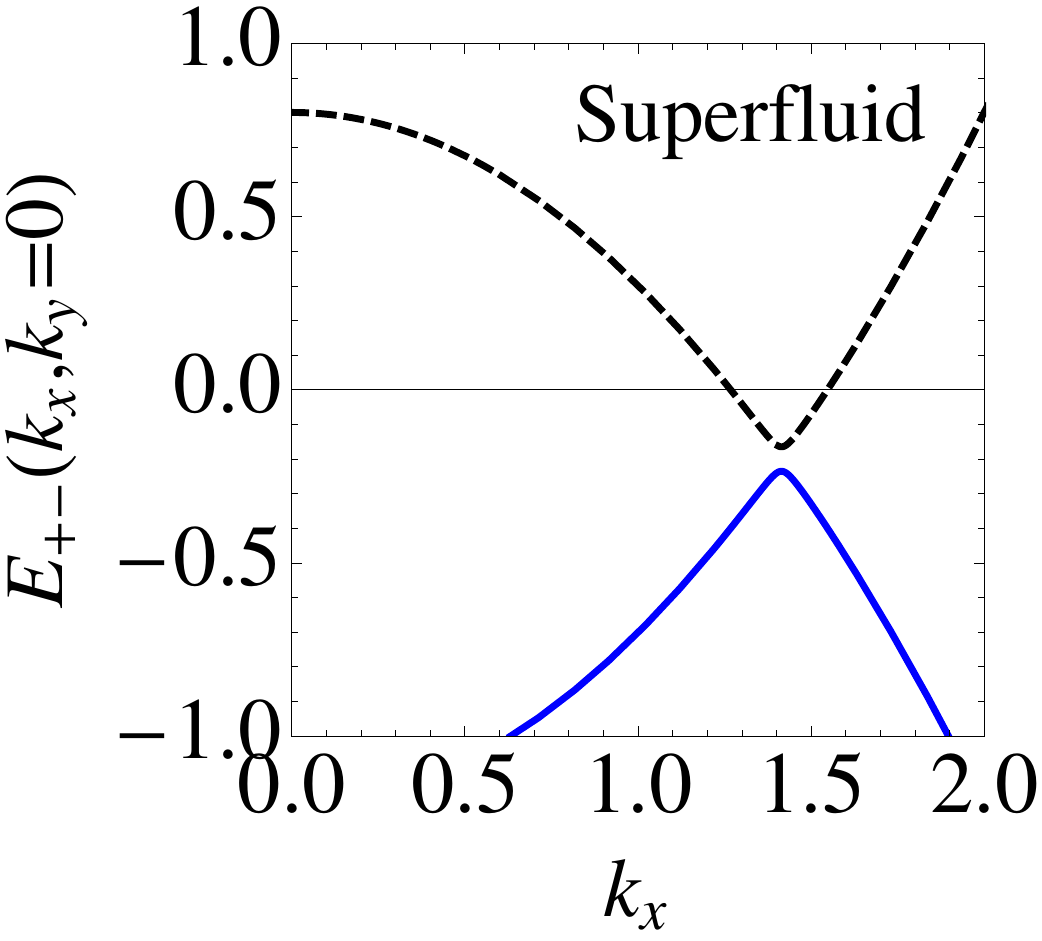}
\vspace*{1mm}
\includegraphics*[width=39mm,angle=0]{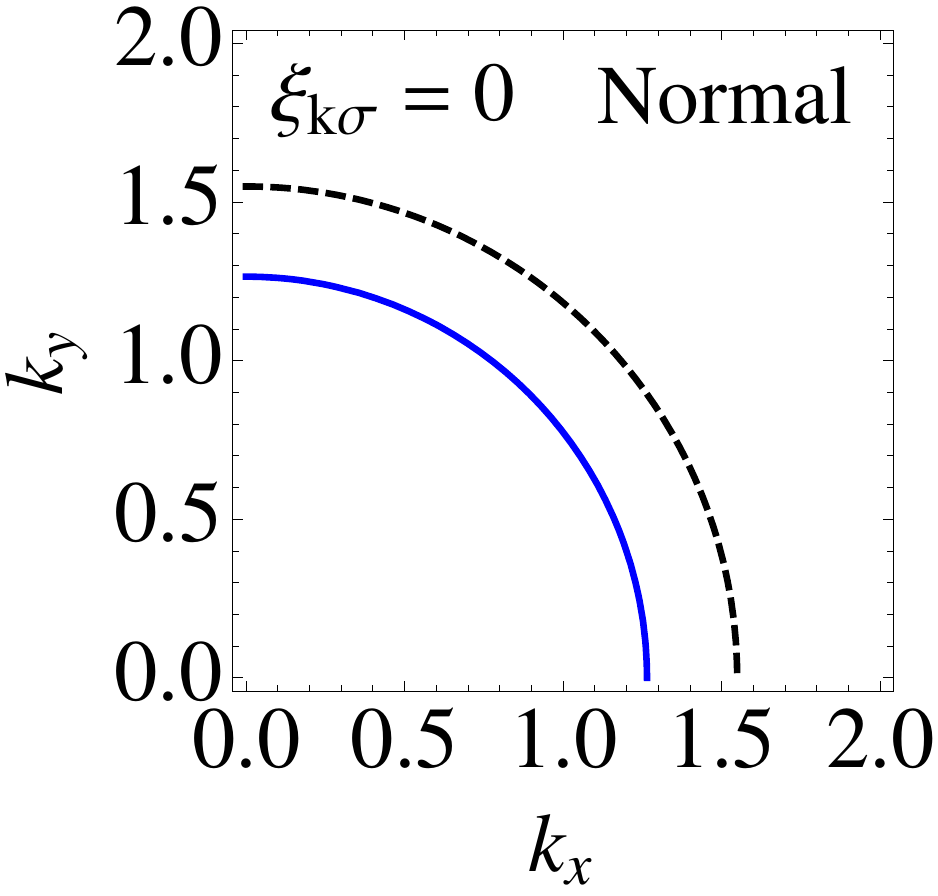}
\vspace*{-2mm}
\caption{(Color online) Superfluid: Meanfield quasi-particle energies in the superfluid phase (left). 
Normal: Mismatched Fermi surfaces in the normal phase (right). The smaller, blue Fermi surface 
is taken to be the $\downarrow$-component, the larger, black-dashed Fermi surface 
belongs to the $\uparrow$-component. 
Here and in the rest of the paper 
we use $\mu=\frac{\mu_{\uparrow}+\mu_{\downarrow}}{2}$ as the 
average chemical potential and $h= \frac{\mu_{\uparrow}-\mu_{\downarrow}}{2}$ as the average 
``Zeeman'' field. We further define the mass ratio as
$r=\frac{m_\downarrow}{m_\uparrow}$. The Fermi momenta of the two species are given by
$k_{\rm F,\uparrow}=\sqrt{2\left(\mu+h\right)}$ and 
$k_{\rm F,\downarrow}=\sqrt{2r\left(\mu-h\right)}$.
Here we used $r = 1$ in units of $m_{\uparrow}$ as appropriate for a equal-mass mixtures, 
$\mu=1$, $h=0.1$, and $\alpha = 0.05$. The superfluid ``gap'' $\alpha$ opens away from the 
Fermi surfaces \cite{sarma63,liu03}
(where the black-dashed line $E_{+}$ crosses zero).
}
\label{fig:FS}
\end{figure}

A mean-field diagonalization of Eq.~(\ref{eq:hamiltonian}) around the static and homogeneous 
projection of superfluid fermion bilinear containing one fermion from each spin species, 
$\alpha = \lim_{q\rightarrow 0} \int_k \psi_{k+\frac{q}{2}\uparrow}\psi_{-k+\frac{q}{2}\downarrow}$, produces 
two quasi-particle branches in the superfluid phase
\begin{align}
E_{\pm} = \frac{\xi_{\mathbf{k}\uparrow}- \xi_{-\mathbf{k}\downarrow}}{2}
\pm
\sqrt{\frac{\alpha^2}{2} +\left(\frac{\xi_{-\mathbf{k}\downarrow} + \xi_{\mathbf{k}\uparrow}}{2}\right)^2}\;,
\label{eq:E}
\end{align}
where the condition $\xi_{\mathbf{k}\sigma} = 0$ 
defines the (bare) Fermi surfaces of the two species in the normal phase. 
This quasi-particle energies can have zeros at finite $\alpha$, as shown in Fig.~\ref{fig:FS}, such 
that the superfluid state can have Fermi surfaces for certain parameters.

Such an exotic superfluid would be an intriguing state of matter: translated to 
the correlated electron context, it would be a 'perfect metal' with large 
Fermi surfaces and a superconductor at the same time. This is in stark 
contrast to both, conventional (s-wave) BCS-superconductors and
high-temperature superconductors: in conventional superconductors, 
the superfluid order parameter gaps out the entire Fermi surface and charge transport occurs 
exclusively via bosonic Cooper pairs; in the high-temperature superconductors, 
the gap has a d-wave structure leaving nodes on the Fermi surface ungapped, however with energetically 
suppressed density of states. 

In the Sarma-Liu-Wilczek superfluid, the simultaneous presence of strong phase fluctuations, that is, the Goldstone mode 
and the gapless fermions makes this a strongly correlated "soup" with anomalous 
transport thermodynamic properties that are not fully understood. Endowing 
such a state with momentum structure in the gap or lattice potentials would 
complement efforts on high-temperature superconductors where 
exotic states of matter with (sometime fractionalized) Fermi surfaces are coupled to 
gapless degrees of freedoms such as gauge bosons.

Parish {\it et al.} \cite{parish07} presented a comprehensive mean-field picture 
of the imbalanced Fermi gas problem at finite temperatures (focusing on the case of dimensionality $d=3$). In particular, they emphasized
the possibility of suppressing the temperature of the tricritical point \cite{sheehy09} by increasing the mass-imbalance 
between the two fermion species. However, it is rather clear that the picture of Refs.~\onlinecite{parish07,sheehy09} 
is not fully realistic in two spatial dimensions, $d=2$. The first- (or second-) order transition 
to the superfluid state at $T>0$ implies the existence of long-ranged 
order and therefore contradicts the Mermin-Wagner theorem. The important role of fluctuations is also 
clearly signaled in the 
recent study \cite{torma13}, operating at Gaussian level.
A complete theory accounting for fluctuations 
should instead (most likely) yield a Berezinski-Kosterlitz-Thouless (BKT) type superfluid involving no true long-ranged order. 
It is on the other hand well 
known that the BKT transition is actually of infinite order, so that the free energy is of the $C^{\infty}$ class, but not an 
analytic function. The occurrence of a tricritical point in the phase diagram 
is therefore not admissible, except at $T=0$.

The bottom line of the previous work was that, in the relatively weakly interacting BCS regime 
of Eq.~(\ref{eq:hamiltonian}), 
mismatching the Fermi surfaces typically lead to abrupt, 
first-order Clogston transitions \cite{clogston62} out of the superfluid state into a phase-separated 
(in real space) state, containing superfluid and normal puddles, which possessed the lowest 
energy among several candidate states 
\cite{yip03,bedaque03,liu04,parish07,pilati08}.
However, most of these analysis have relied on mean-field theory or focussed on 
three dimensional Fermi mixtures.

It is worthwhile noting that fluctuation effects are known to change the basic properties of phase transitions in a number of important systems, in low dimensionalities in particular. It seems that more commonly fluctuations favor the first-order scenario. For a review on the classical sytems we refer to \cite{Binder87}. On the quantum side, we mention the fluctuation-induced first-order quantum phase transitions in itinerant magnets and superconductors \cite{Lohneysen07, Belitz05} and the quantum variant of the Imry-Ma phenomenon  recently established rigorously \cite{Greenblatt09}. This was also firmly recognised that a wide class of metallic phase transitions is first order at low temperatures, but (by necessity) second-order at sufficiently high $T$ \cite{Jakubczyk10}.   

\section{Key results}
\label{sec:key}
%
\begin{figure}[t]
\vspace*{1mm}
\includegraphics*[width=90mm,angle=0]{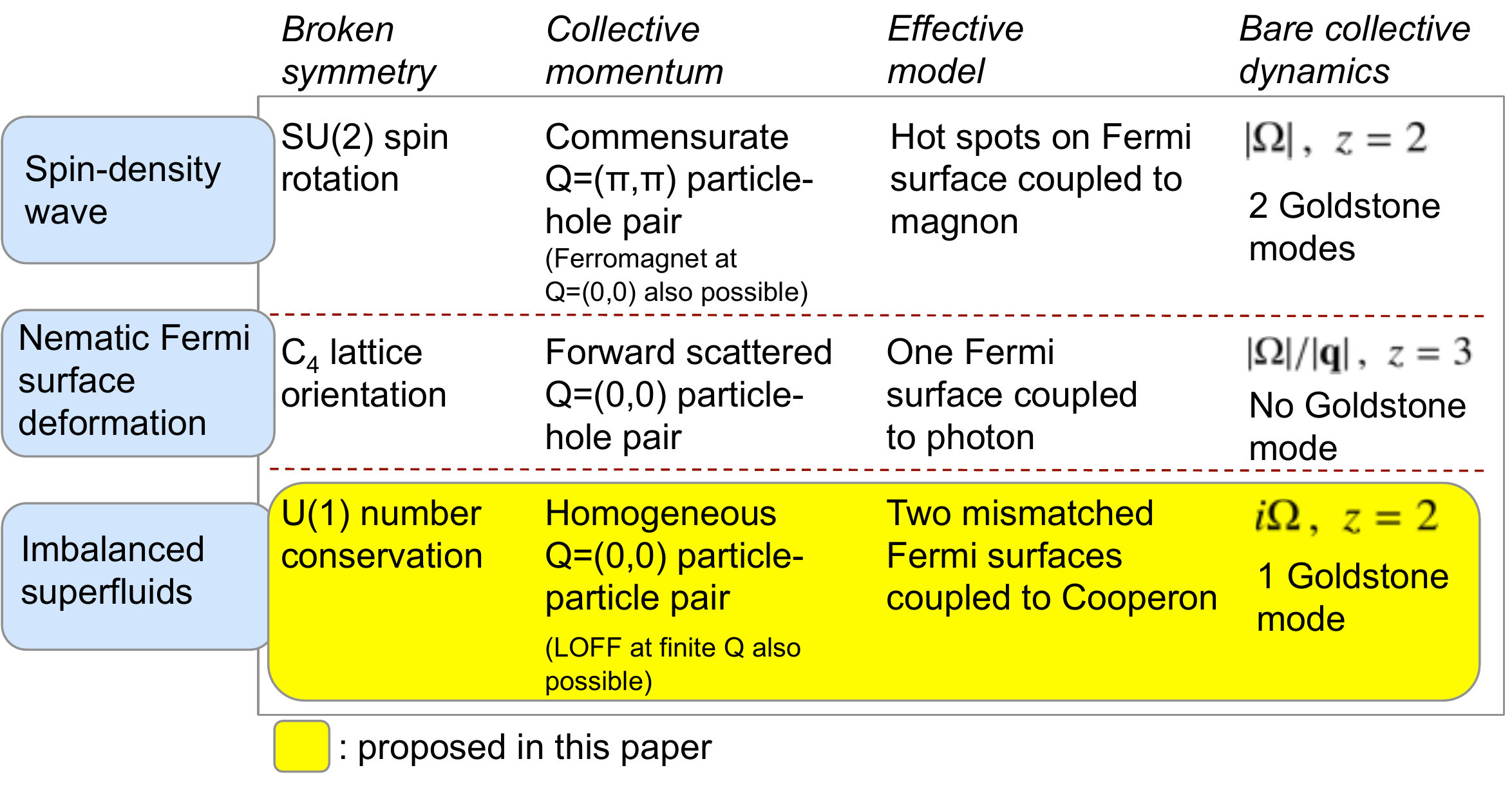}
\caption{(Color online) Plain-vanilla non-Fermi liquid "metal" quantum phase 
transitions at finite fermion density in two spatial dimensions. This plain-vanilla 
classification excludes exotic scenarios such as e.g.: topological order, 
fractionalization/emergent gauge field or transitions to fully insulating states. Non-Fermi liquid properties of 
the spin-density wave and Fermi surface deformation cases have recently been discussed 
for example in Ref.~\onlinecite{max10,jun12,abrahams13,kivelson01,metzner03,lee09}.
The interplay with spatially inhomogenous Larkin-Ovchinikov-Fulde-Ferell (LOFF) states 
is discussed in the text.}
\label{fig:table}
\end{figure}
\begin{figure}
\vspace*{-8mm}
\hspace*{-7mm}
\includegraphics*[width=102mm,angle=0]{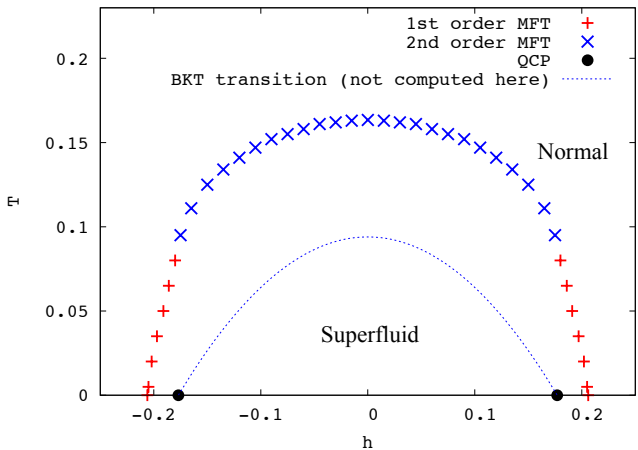}
\vspace*{-10mm}
\caption{(Color online) Fluctuation-corrected phase diagram of imbalanced fermionic superfluids in two 
dimensions. Goldstone and longitudinal fluctuations induce new quantum critical points (black circles) at $(h=\pm 0.18, T = 0)$ 
between a normal Fermi mixture 
and a Sarma-Liu-Wilczek superfluid with gapless fermions. Toward $h\rightarrow0$, another transition to 
a fully gapped superfluid is expected. Furthermore fluctuations suppress 
spontaneous symmetry breaking to $T=0$. Extension
of our computations to finite $T$ \cite{gersdorff01} would lead to a BKT transition line connecting the 
new QCPs \cite{bkt_footnote}. Mean-field theory (MFT) predicts an incomplete picture: a superfluid dome (blue crosses) 
with long-range order and 1st order transitions on its wings (red crosses). 
}
\label{fig:domes}
\end{figure}

This paper addresses the impact of order parameter fluctuations 
on imbalanced superfluids in interacting Fermi mixtures in two dimensions. 
We propose here the possibility of a new type of fluctuation-induced quantum criticality 
between imbalanced superfluids and a normal Fermi gas as realizable with present-day 
experimental technology in the relatively weakly interacting BCS regime. 
We have written this paper hoping that the fluctuation-induced 
quantum criticality of superfluid, charge-neutral Fermi mixtures will become a fruitful addition to the 
``plain vanilla" set of non-Fermi liquid quantum critical points at finite fermion density tabulated in 
Fig.~\ref{fig:table}. To substantiate this claim, we here pursue two complementary 
directions: 

In Sec.~\ref{sec:flow}, we present first computational evidence that 
strong phase and amplitude fluctuations of the superfluid order parameter 
suppress the regime of phase-separation and lead to a continuous quantum phase transition  
between an imbalanced Sarma-Liu-Wilczek superfluid with gapless Fermi surfaces 
and a normal mixture as shown in Fig.~\ref{fig:domes}. This phase diagram is 
based on the solution of a 
partial differential (renormalization group flow) equation for the effective potential/free energy density of the Fermi mixture 
$U\left(\alpha\right)$. 
The possibility of a quantum phase transition from 
a state of ``2 Fermi surfaces + BEC'' to "2 Fermi surfaces and no BEC" 
was alluded to in the different microscopic context of mixtures of fermions with fundamental bosons \cite{powell05}. 
Because of the fundamental nature of the boson in Ref.~\onlinecite{powell05}, this transition 
does not fall into the plain-vanilla non-Fermi liquid ``metal'' categorization of Fig.~\ref{fig:table}, 
which restricts to purely fermionic systems.

\begin{figure}
\includegraphics*[width=80mm,angle=0]{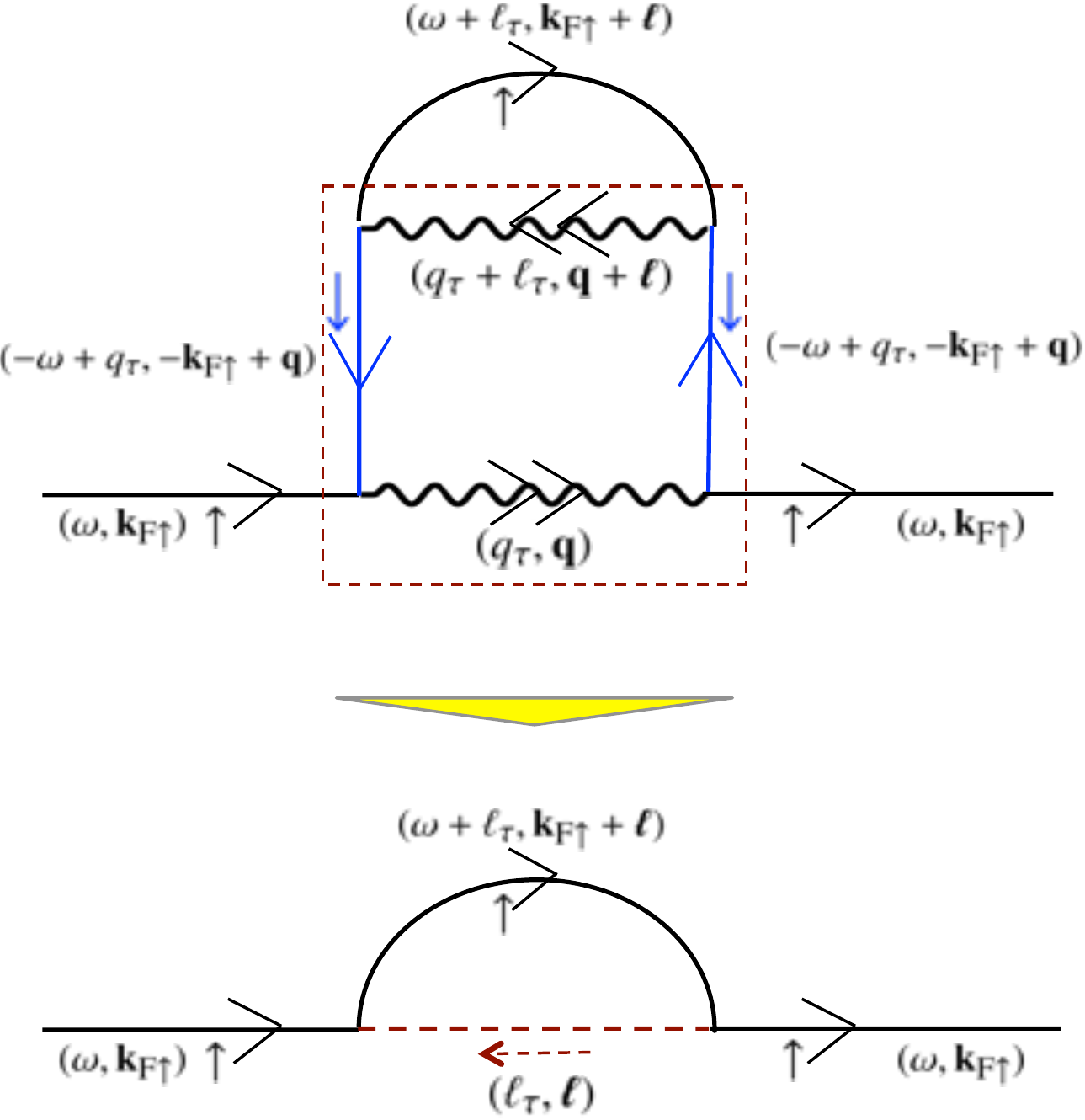}
\vspace*{-2mm}
\caption{
(Color online) Repeated scattering of Cooperons and particle-hole 
pairs of cold, off-shell fermions (red, dashed box in upper graph) 
mediates a singular, intra-species 
interaction (red, dashed line in bottom graph) that induces 
non-Fermi liquid behavior on the Fermi surfaces. 
Wiggly line here denotes the one-loop dressed (cf. Fig.~\ref{fig:ppbubble}) 
Cooper pair propagator.
}
\label{fig:diagrams}
\end{figure}

In Sec.~\ref{sec:diagrammatics}, we analyze the diagrammatics of the effective field theory coupling 
the collective ``Cooperon'' to two mismatched Fermi surfaces. Non-Fermi liquid physics appears at 
two-loop order, despite the fact that the Cooperon reconstructs the fermion dispersions away from the 
Fermi surfaces. Behind this are repeated scattering processes involving Cooperons and off-shell particle-hole pairs of 
the respective {\it other} species as shown in Fig.~\ref{fig:diagrams}. This enhances the scattering rate around 
the Fermi surface which should be observable with experimental techniques such as 
rf-spectroscopy \cite{stewart08, grimm12}. One may therefore say, on mean-field level, both Fermi surfaces 
remain ``cold'' at the superfluid transition. Beyond mean-field, however, order parameter fluctuations 
``warm them up'' leading to faster than Fermi liquid decay rates but probably 
ultimately stable quasi-particles. Different, interesting mechanisms to 
achieve non-Fermi liquid features with imbalanced Fermi gases were given in Refs.~\onlinecite{lamacraft10,lan13}. 
Putting the attractively interacting  Fermi mixture into spin-dependent optical lattices \cite{fisher09}, that deform
the bare Fermi surfaces such that ``band-crossing'' are generated, also naturally leads to non-Fermi liquid physics 
from pairing at least close to the ``hot'' crossing points.

Our results of Secs.~\ref{sec:flow},\ref{sec:diagrammatics} undoubtedly point out the 
qualitative importance of fluctuations to the physics of imbalanced superfluids in two dimensions. 
Like any other 
quantum critical point for interacting Fermi systems at finite density, however, fluctuations in the 
vicinity of the QCPs (black dots in Fig.~\ref{fig:domes}) will attract competition from other instabilities. 
To this end, two important aspects should be investigated further.

The first one is the 
interplay with induced intra-species p-wave superfluidity \cite{schwenk06,sheehy11}, whose ordering scales, 
however, are expected to be very small in the relatively weakly interacting BCS regime. 
In fact, there are similarities to the case of induced {\it d-wave pairing} close to antiferromagnetic critical points 
in metals \cite{max10,hartnoll11}. There, singular magnetic fluctuations have two counter-acting effects:
they reduce the amount of electronic spectral weight around the Fermi surfaces (especially the hot spots), 
and at the same time they generate logarithmic singularities in the d-wave pairing channel.
A similar competition occurs here in imbalanced superfluids: 
singular Cooper pair fluctuations generate intra-species pairing tendencies in the p-wave channel; at the same time 
the lifetimes of quasiparticles around the Fermi surfaces are reduced by the mechanism of Fig.~\ref{fig:diagrams}. 

The second open question is the interplay with translational symmetry breaking tendencies from 
inhomogeneous LOFF pairing (see for example \cite{hulet10,lamacraft10,scalettar12,roscher13} 
for analysis in one-dimensional systems). As pointed out by Lamacraft \cite{lamacraft10}, 
scattering off singular Cooperons with spatial modulation is a natural route to non-Fermi liquid behavior
because it directly connects the mismatched Fermi surfaces. At the mean-field level, Sheehy and Radzihovsky 
provided a comprehensive analysis of the imbalanced Fermi gas in the continuum \cite{sheehy07} with the 
result that the inhomogeneous LOFF states occupies only a very narrow strip in the phase diagram. A very recent study \cite{torma13}
emphasises the prominent role of fluctuations in $d=2$ and suggests that (even at the Gaussian level) the LOFF states are unstable at $T>0$. 

In lower dimensionality $d<2$ and under inclusion of lattice potentials, periodic modulations of the 
phase and or the amplitude of the superfluid order parameter can be stabilized at the mean-field 
level \cite{lutchyn11} and an extension of our theory to this case is an interesting future direction. 
A fluctuation theory beyond Gaussian level for this case 
is significantly more complicated: in addition to the global (number) charge symmetry, 
also continuous translation symmetry is broken leading to an additional Goldstone ``phonon'' in the 
fluctuation spectrum. One would then deal with a non-Fermi liquid coupled to a damped, 
bosonic sector of the $O(4)$ universality class.

We conclude in Sec.~\ref{sec:conclusion} and give a brief outlook on experimental observation with ultracold 
quantum gases.

\section{Functional Flow to criticality}
\label{sec:flow}

For interacting Fermi systems at the brink of phase-separating and abrupt jumps of the order parameter 
(as is the case in the mean-field theory for imbalanced Fermi mixtures at low temperatures), an expansion 
of the effective potential in powers of the order parameter is not possible. 
The approach Ref.~\cite{pawel09} proposed to integrate a {\it functional} flow equation of the local potential 
along a continuous flow parameter $\Lambda$ this way capturing order parameter fluctuations keeping 
the {\it entire} local potential on a discrete set of points in field space. The initial conditions for this flow 
at the scale $\Lambda_0$ (the characteristic energy below which collective order parameter fluctuations 
become important) is given by the mean-field effective potential of Eq.~(\ref{eq:hamiltonian}) 
\begin{align}
U^{\Lambda=\Lambda_0}(\alpha)=\frac{-1}{2 g}\alpha^2
-
T\sum_{k_0,\mathbf{k}}
\ln\left[
\frac{
\left(-i k_0 + \xi_{\mathbf{k}\uparrow}\right)
\left(i k_0 + \xi_{-\mathbf{k}\downarrow}\right)
+\alpha^2/2}
{
\left(-i k_0 + \xi_{\mathbf{k}\uparrow}\right)
\left(i k_0 + \xi_{-\mathbf{k}\downarrow}\right)
}
\right]\;,
\label{eq:local_pot}
\end{align}
where $\alpha$ is the real-valued ordering field (the longitudinal 
$\sigma$-direction in the local, mexican hat potential that involves all 
powers of the ordering field). The mean-field theory is derived in  
Appendix {\ref{app:mft}}. The transverse Goldstone quantum field will be denoted 
by $\pi$. In two dimensions, the attractive interaction
%
$g(\Lambda_{\mathbf{k}_{\rm F}})=-\frac{2\pi}{m \ln\left[a_{\text{2D}}\Lambda_{\mathbf{k}_{\rm F}}\right]}$
%
changes logarithmically with momentum scale \cite{bertaina11,langmack12} and we will 
evaluate it at a characteristic scale of the order of the Fermi momenta involved, $\Lambda_{\mathbf{k}_{\rm F}}$.
The particle-particle ladder for two particles in vacuum has a logarithmic 
singularity for small and large momenta. Therefore, two particles form a bound state for arbitrarily weak 
attraction among them \cite{bloom75,randeria89}. 
Here $a_{\text{2D}}$ is the 2D scattering length. 
This paper restricts to the BCS regime, 
$|\Lambda_{\mathbf{k}_{F}}|a_{\text{2D}}\gg1$ such that gaps are relatively 
small and we are well separated from the strong-coupling region 
where the crossover to the BEC regime sets in \cite{bertaina11}. 
We may therefore expect BCS-BEC crossover-like contractions of the Fermi volumes 
\cite{leggett80} to be subdominant and we perform our 
calculations at constant chemical potentials (using their difference as the tuning parameter).

\begin{figure}
\includegraphics*[width=55mm,angle=0]{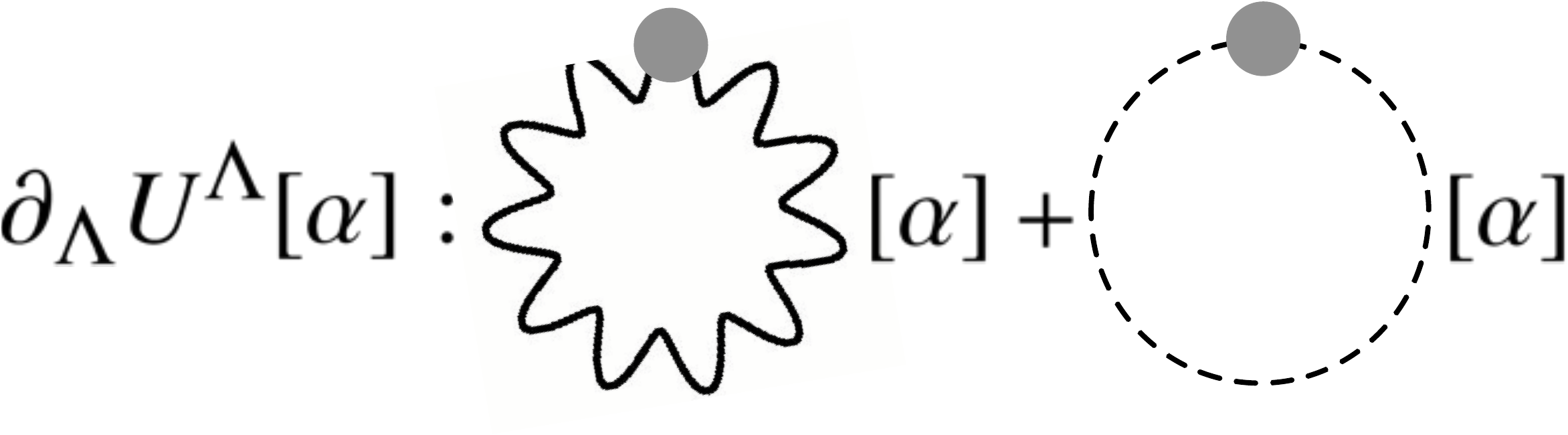}
\caption{
(Color online) Graphical representation 
of the flow equation Eq.~(\ref{eq:flow_U}) involving traces 
over field ($\alpha$)-dependent Goldstone (wiggly loop) and longitudinal (dashed loop) propagators, 
respectively. Grey dots denote cutoff insertions.
}
\label{fig:flow}
\end{figure}

We now proceed to integrate order parameter fluctuations with the functional flow 
equation for the effective potential
\begin{align}
\partial_{\Lambda}U^{\Lambda}(\alpha)=\frac{1}{2}\text{tr}_q
\left[
\partial_\Lambda R^{\Lambda}(q)\left(
G^{R}_{\sigma\sigma}(q;\alpha)
+
G^{R}_{\pi\pi}(q;\alpha)
\right)
\right]\;.
\label{eq:flow_U}
\end{align}
where 
$\text{tr}_{q}=T\sum_{q_\tau}\int \frac{d^2\mathbf{q}}{(2\pi)^2}$ stands for frequency and momentum 
integrations over the field-dependent Goldstone and longitudinal loops shown in Fig.~\ref{fig:flow}. 
At vanishing cutoff scale $\Lambda\to 0$ all the fluctuation modes are incorporated into 
the partition function and $F(\alpha) = \lim_{\Lambda\to 0}U^{\Lambda}(\alpha)$ is the free 
energy density. Both, the longitudinal propagator $G^{R}_{\sigma\sigma}(k;\alpha)$ and the Goldstone 
propagator $G^{R}_{\pi\pi}(k;\alpha)$ are explicitly field ($\alpha$)-dependent
\begin{align}
G^R_{\pi\pi}(q;\alpha)& = \frac{Z_{\Omega}(\alpha_0) q_\tau^2+ Z_{\mathbf{q}}(\alpha_0)\mathbf{q}^2  
+
m^2_{\pi\pi}(\alpha,R)
}
{\text{det}^{R}(q;\alpha)}
\label{eq:matrix_element_props}\\
G^R_{\sigma\sigma}(q;\alpha) &= \frac{Z_{\Omega}(\alpha_0) q_\tau^2+ Z_{\mathbf{q}}(\alpha_0)\mathbf{q}^2  
+ 
m^2_{\sigma\sigma}(\alpha,R)  
}
{\text{det}^{R}(q;\alpha)}\;.
\nonumber
\end{align}
The above expressions involve an approximation, treating the $Z$-factors as constants (see below). 
The ``mass'' terms for the Goldstone boson
$m^2_{\pi,R}(\alpha)=U'(\alpha)+R^\Lambda(q)$ and 
the longitudinal propagator 
$m^2_{\sigma,R}(\alpha)=U'(\alpha)+\alpha^2 U''(\alpha)+R^\Lambda(q)$ 
involve a regulator $R^\Lambda(q)$, to be specified below, and are 
evaluated at general $\alpha$. Here, $U'(\alpha)$ is a first, and $U''(\alpha)$ is the second 
field derivative with respect to $\rho = \frac{1}{2}\alpha^2$. The determinant in $(\sigma,\pi)$ field space 
is 
%
\begin{align}
\text{det}^{R}(q;\alpha) = &
\left(Z_{\Omega}(\alpha_0) q_\tau^2+ Z_{\mathbf{q}}(\alpha_0) \mathbf{q}^2 + m^2_{\sigma\sigma}(\alpha,R) \right)
\times
\label{eq:det}
\\
&\left(Z_{\Omega}(\alpha_0)  q_\tau^2+ Z_{\mathbf{q}}(\alpha_0) \mathbf{q}^2 + m^2_{\pi\pi}(\alpha,R) \right)
+
X^2(\alpha_0)  q_\tau^2\;.
\nonumber
\end{align}

The $Z$- and $X$-factors are frequency and momentum renormalization factors and are 
evaluated at the minimum of the effective potential $\alpha_0$. In the superfluid phase away from the 
critical point, $\sigma$ and $\pi$ components of the $Z$-factors can exhibit very different scaling properties 
in the infrared \cite{pistolesi04, strack08}. Since our 
main interest here is the phase diagram and not the infrared asymptotics, 
we will keep here only one set of $Z$- and $X$-factors. The $X$-factor 
multiplies a linear time-derivative  $G^R_{\sigma\pi} \sim X q_\tau$ and couples the Goldstone and 
longitudinal fluctuations. In the present renormalization  scheme we supplement only the diagonal components of the inverse propagator with the cutoff term $R$. Within such a procedure, there arises no contribution from  $G^R_{\sigma\pi}$ in Eq.~(\ref{eq:flow_U}) and the impact of the $X$-factor on the flow of $U$ occurs only via the determinant. 

Eq.~(\ref{eq:flow_U}) emerges upon evaluating the exact flow equation for the generator of 1-particle-irreducible vertices \cite{wetterich93} at a uniform field configuration. The resultant formula involves field- and scale- dependent $Z$ and $X$ factors. The essence of the approximation leading to the closed set Eq.~(\ref{eq:flow_U}-\ref{eq:det}) amounts to disregarding these dependencies. 
An important ingredient of the present study resides in the local potential Eq.~(\ref{eq:local_pot}), which is the starting point of for the solution of Eq.~(\ref{eq:flow_U}), and which is directly and clearly related to the microscopic, fermionic system 
(see Appendix \ref{app:mft} for a derivation).


Let us also note that Eq.~(\ref{eq:flow_U}-\ref{eq:det}) fulfill the Mermin-Wagner 
theorem \cite{berges02} upon turning on a temperature, and provide a 
framework to study quantum fluctuations beyond BCS-theory in a superfluid state 
{\it without} violating Goldstone's theorem, that is, without generating an 
unphysical Goldstone mass \cite{diener08,strack08,obert13}. This is because the flow of the minimum 
of the potential $\alpha_0$ is self-consistently determined from 
the condition (Ward identity) that the (unregulated) Goldstone mass 
$m_{\pi,R=0}[\alpha_0]^2=U'[\alpha]\equiv0$ vanishes for all $\Lambda$.


%
\begin{figure}[t]
\vspace*{1mm}
\includegraphics*[width=75mm,angle=0]{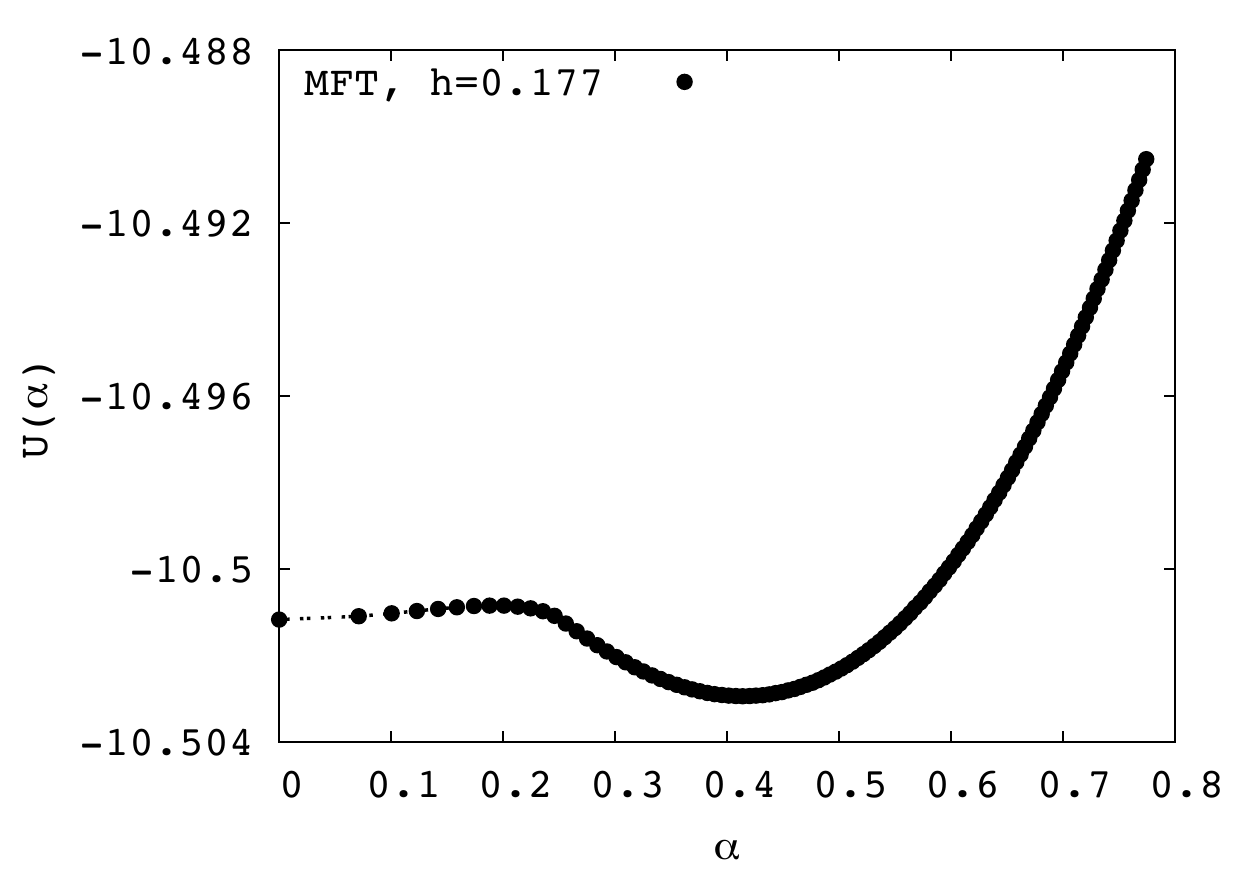}
\includegraphics*[width=75mm,angle=0]{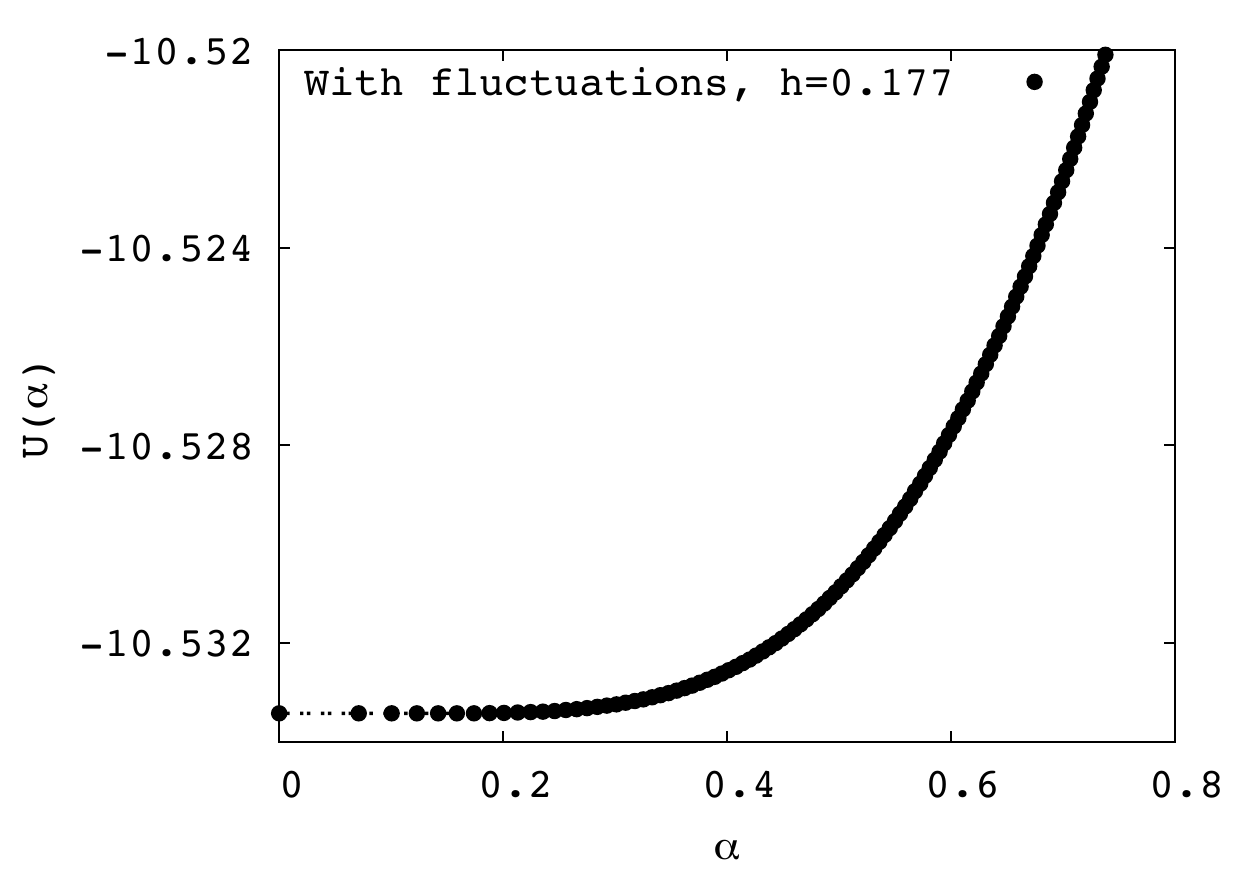}
\caption{(Color online) Comparison of the zero-temperature potentials at the new QCP (black circles 
at $h=0.177$, T=0 in Fig.~\ref{fig:domes}). Mean-field theory (top) incorrectly predicts a phase-separated fully gapped BCS + excess fermion 
state with the global minimum at $\alpha_0 =0.42$, which overshadows the Fermi surface mismatch.
Fluctuations (bottom) smear out the curved region in the potential, and in fact 
lead to a quantum phase transition with continuously onsetting superfluid order $\alpha_0$ slightly above zero.
Numerical parameters and initial values of the $Z-$ and $X$-factors are:
$g=-2$, $Z^{\Lambda_0}_{\Omega}=Z^{\Lambda_0}_{\mathbf{q}}=1$, and $X^{\Lambda_0}=0.2$.
Further technical details are given in Appendix \ref{app:details}.
}
\label{fig:h0177}
\end{figure}
\begin{figure}[t]
\vspace*{1mm}
\includegraphics*[width=75mm,angle=0]{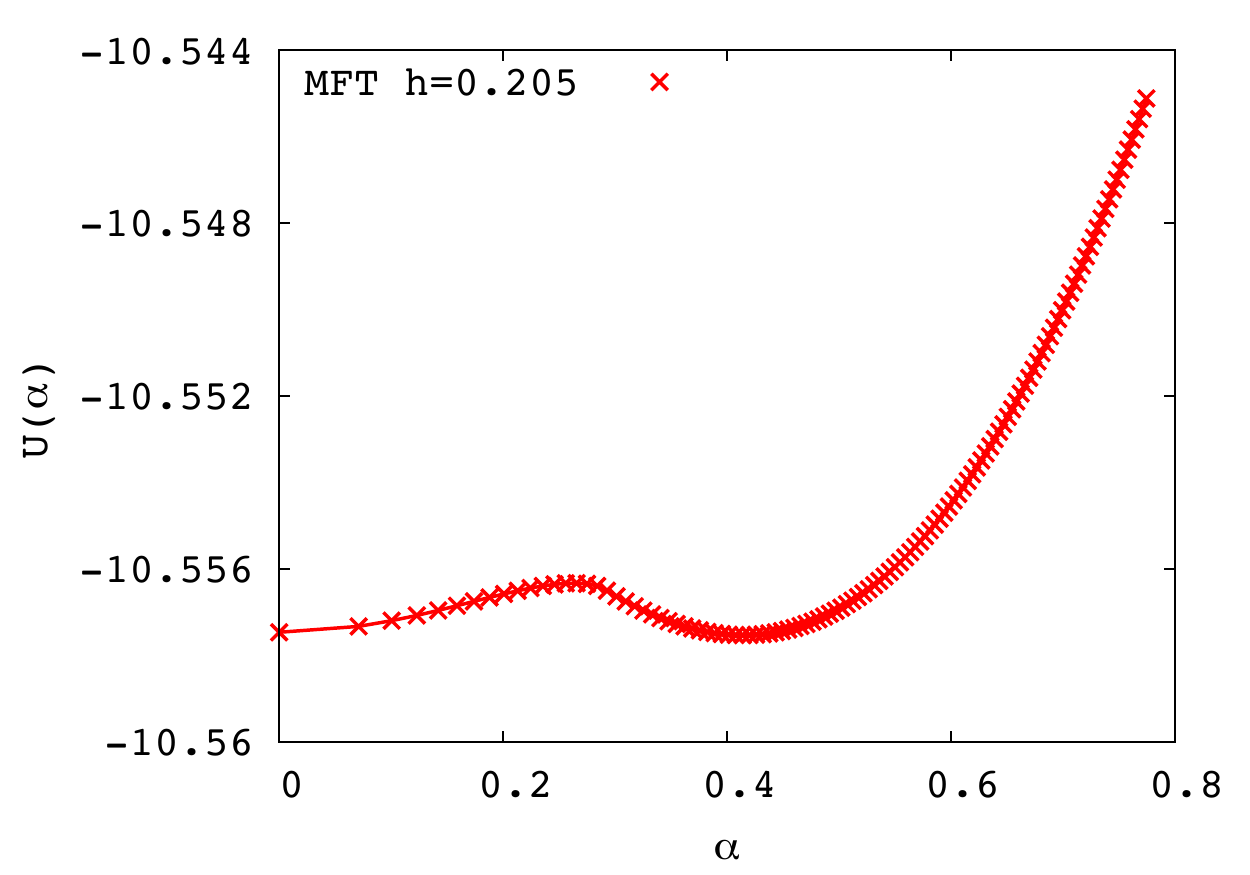}
\includegraphics*[width=75mm,angle=0]{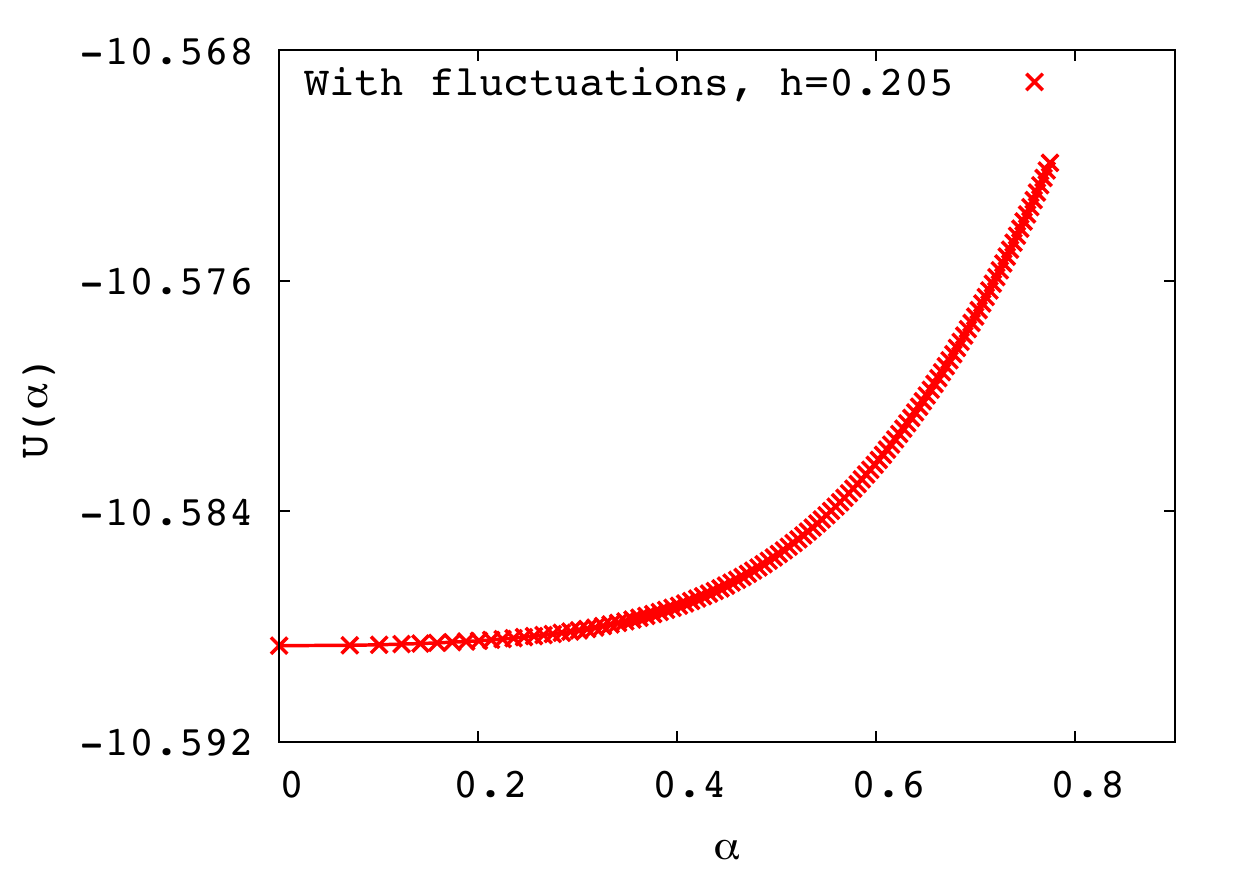}
\caption{(Color online) Comparison of the zero-temperature potentials at the putative 
mean-field first order transition (red cross at $h=0.205$, $T=0$ in Fig.~\ref{fig:domes}). Mean-field theory (top) incorrectly predicts a 
first-order transition to a phase-separated fully gapped BCS + excess fermion 
state with a second minimum at $\alpha_0 =0.41$ degenerate to the $\alpha_0=0$ one, the former overshadowing 
the Fermi surface mismatch.
Fluctuations (bottom), however, suppress the ordering tendencies and instead 
yield a symmetric state with $\alpha_0=0$. The other initial parameters are the same 
as in Fig.~\ref{fig:h0177}. }
\label{fig:h0205}
\end{figure}

We now present our results from solving the functional flow of Eq.~(\ref{eq:flow_U}-\ref{eq:det}). 
Using a combined frequency and momentum cutoff 
%
$R_\Lambda (q_\tau,\mathbf{q}) = Z_{\mathbf{q}}\left(\Lambda^2-q^2-\frac{Z_\Omega}{Z_{\mathbf{q}}}q_\tau^2\right)\theta \left(\Lambda^2-q^2-\frac{Z_\Omega}{Z_{\mathbf{q}}}q_\tau^2\right)$
%
the frequency and momentum integrations can be performed and Eq.~(\ref{eq:flow_U}) becomes
%
\begin{align}
\Lambda\partial_\Lambda  U_\Lambda (\rho) =& 
-\frac{\tilde{Z}\Lambda^2}{4\pi^2}\left(M_{\sigma\sigma}(\rho)+M_{\pi\pi}(\rho)\right)
\times
\Bigg\{
-\frac{\tilde{Z}\Lambda}{X^2} 
+
\nonumber\\
&\left(
\frac{\Lambda^2}{X\sqrt{M_{\pi\pi}(\rho) M_{\sigma\sigma}(\rho)}} 
+\frac{\tilde{Z}^2\sqrt{M_{\pi\pi}(\rho) M_{\sigma\sigma}(\rho)}}{X^3}
\right)
\nonumber\\
&
\arctan\left[\frac{\Lambda X}{\tilde{Z}\sqrt{M_{\pi\pi}(\rho) M_{\sigma\sigma}(\rho)}}\right]
\Bigg\} \;.
\label{eq:mouse_hole}
\end{align}
%
with $\rho = \alpha^2/2$, $\tilde{Z}\equiv \sqrt{\frac{Z_\Omega}{Z_{\mathbf{q}}}}$,
the regulated longitudinal mass $M_{\sigma\sigma}(\rho) \equiv Z_{\mathbf{q}}\Lambda^2 +U'(\rho)+2\rho U''(\rho)$ and 
the regulated Goldstone mass $M_{\pi\pi}(\rho) \equiv Z_{\mathbf{q}}\Lambda^2 +U'(\rho)$.

Eq.~(\ref{eq:mouse_hole}) is a second-order partial differential equation, which we 
solve numerically on a grid in field space containing up to 200 points, where the non-zero minimum (initial) is situated around the middle of the box. The initial condition 
for $U(\alpha)$ is the un-truncated, mean-field effective potential containing all orders 
in the order parameter field. 

As described in Fig.~\ref{fig:domes}, fluctuations may have a drastic effect
on the even qualitative features of the phase diagram. Here, quantum fluctuations suppress the first-order transition 
and associated phase separation between a BCS + excess fermion 
state and instead indicate a continuous quantum phase transition to a superfluid state with two Fermi surfaces, 
which could exist as an intermediate phase. Upon further increasing the interaction or 
reducing the imbalance $h$, this is expected to become a fully gapped BCS state with no Fermi surface transiting over a state 
one Fermi surface with enhanced density of states (when the black-dashed quasi-particle branch in Fig.~\ref{fig:FS} is pushed 
upwards by larger values of the gap $\alpha > h$) \cite{liu04}. Note that the most significant renormalization of the potential $U(\rho)$ occurs relatively early in the flow (intermediate $\Lambda$), where the key effect is that regions with large curvature are smoothened out by quantum 
fluctuations. Figs.~\ref{fig:h0177} and \ref{fig:h0205} exhibit a comparison of the mean-field potential with 
its renormalized version at various points in the phase diagram. 

Let us observe that the above situation does not occur for sufficiently large absolute values of the coupling $g$, in which  case fluctuations are not sufficiently potent to induce a QCP and the transition at $T=0$ remains first-order also after renormalization. By tuning $g$ one goes between the two possibilities crossing the specific value of $g$, where the system displays a quantum tricritical point.

We also note here the reassuring feature of Eqs.~(\ref{eq:flow_U},\ref{eq:mouse_hole}) 
 that at any finite $T$, a solution with a finite minimum $\alpha_0$, which would indicate spontaneous symmetry-breaking 
is not permissible. This is because of a logarithmic running of $\alpha_0$ from the Goldstone loop such that 
$\alpha_0 \rightarrow 0$ for $\Lambda\rightarrow 0$ at finite $T$.

We close this section by noting that the ultimate fate of the intermediate gapless 
superfluid in two dimensions will also depend strongly on fermionic self-energy effects, 
such as those discussed in the 
next Sec.~\ref{sec:diagrammatics} and fluctuations in the shape 
of the Fermi surfaces, and the interplay with competing instabilities. An intriguing 
scenario \cite{yamase11} is the possibility of a merging of the two QCP's in 
Fig.~\ref{fig:domes} to form an isolated QCP at $(h=0,T=0)$ indicating the absence of any
superfluid phase in the ground state. Resolution of such possibilities requires studying 
the interplay of the bosonic potential and single-particle, fermionic effects.

\section{Non-Fermi liquid Mechanism}
\label{sec:diagrammatics}

We now outline how the single-fermion properties 
are affected by order parameter fluctuations at the 
imbalanced superfluid quantum phase transition. 
To this end, we first write down the effective quantum 
field theory coupling the two mismatched Fermi surfaces 
to the (bosonic) Cooper pairing field containing 
one fermion from each Fermi surface. We then present 
a repeated scattering mechanism that is expected to
reduce the lifetime of the quasiparticles around the Fermi surfaces 
leading to deviations from the Fermi liquid.

\begin{figure}[b]
\vspace*{-2mm}
\includegraphics*[width=60mm,angle=0]{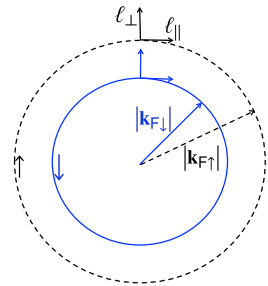}
\vspace*{-2mm}
\caption{
(Color online) Mismatched Fermi surfaces and 
local coordinate systems in momentum space.
}
\label{fig:spheres}
\end{figure}

\begin{figure}[t]
\vspace*{-10mm}
\includegraphics*[width=80mm,angle=0]{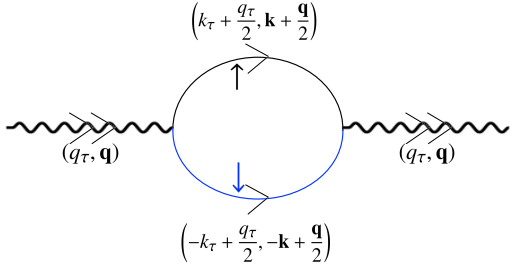}
\vspace*{-10mm}
\caption{
(Color online) Imbalanced particle-particle bubble that 
determines the collective Cooper pair dynamics (the $X$ and $Z$-factors)
given in Eq.~(\ref{eq:phiphi}). Generalized to the superfluid phase, 
fermionic contractions of this form determine the initial values of the 
Goldstone and longitudinal propagtors in the renormalization group flow of 
Eqs.~(\ref{eq:matrix_element_props},\ref{eq:det}). Wiggly line 
is the bosonic Cooper pair propagator, blue straight line the spin-down fermion 
propagators and, black straight line the spin-up fermion propagator.
The cubic vertex connecting the three is the Yukawa coupling.
}
\label{fig:ppbubble}
\end{figure}

The relevant Lagrangian density contains the propagators of 
the two Fermion species
\begin{align}
\mathcal{L}_{\overline{\psi}\psi}=
\overline{\psi}_{k\uparrow}\left(
-ik_\tau + \xi_{\mathbf{k}\uparrow}
\right)\psi_{k\uparrow}
+
\overline{\psi}_{k\downarrow}\left(
-ik_\tau + \xi_{\mathbf{k}\downarrow}
\right)\psi_{k\downarrow}\;,
\end{align}
with $\xi_{\mathbf{k}\sigma}$  
fermion dispersions leading to two mismatched 
Fermi surfaces (Fig.~\ref{fig:spheres}), 
a complex-valued, dynamical Cooper pairing field
\begin{align}
\mathcal{L}_{\phi^\ast\phi}
=\phi^\ast_q
\left(
-X\, i q_\tau +  Z_{\Omega} q_\tau^2 + Z_{\mathbf{q}}\mathbf{q}^2
+r
\right)
\phi_q\;,
\label{eq:phiphi}
\end{align}
whose leading frequency ($q_\tau$) and momentum ($\mathbf{q}$) dependence 
is determined by the particle-particle bubble explained in Fig.~\ref{fig:ppbubble}.
$r$ is the usual control parameter; we will here be focussed on the quantum critical point 
where $r=0$. 
The complex, linear frequency term determines the bare collective dynamics 
to have dynamical exponent $z=2$. Note that the 
Cooperon propagator Eq.~(\ref{eq:phiphi}) is only valid when frequencies 
$q_\tau$ and momenta $\mathbf{q}$ are small when compared to the Fermi energy.
At higher scales, the bi-fermion character of the $\phi$  bosons results 
in a (decay) continuum and branch cuts in the complex plane, as per the logarithm 
of Eq.~(\ref{eq:log}); then the simple pole structure of 
Eq.~(\ref{eq:phiphi}) is insufficient to describe the collective dynamics properly. 
Expanding this logarithm gives the expressions for the $X$- and $Z$-factors
(see Appendix \ref{app:ppbubble}).

The Cooper pairs are coupled to the pair of mismatched Fermi surfaces 
via a Yukawa coupling with strength $g$
\begin{align}
\mathcal{L}_{\psi\phi}
=
g\left(
\phi_q \overline{\psi}_{-k+\frac{q}{2}\uparrow} \overline{\psi}_{k+\frac{q}{2}\downarrow } 
+
\phi_q^\ast
\psi_{k+\frac{q}{2}\downarrow } 
\psi_{-k+\frac{q}{2}\uparrow}
\right)
\label{eq:yukawa}
\end{align}
such that they can decay and recombine into fermions. Ref.~\onlinecite{powell05} 
wrote down related critical field theories for Bose-Fermi mixtures with the aim 
of describing Fermi-Bose molecule formation of the fermions with fundamental bosons.

\begin{figure}[t]
\includegraphics*[width=75mm,angle=0]{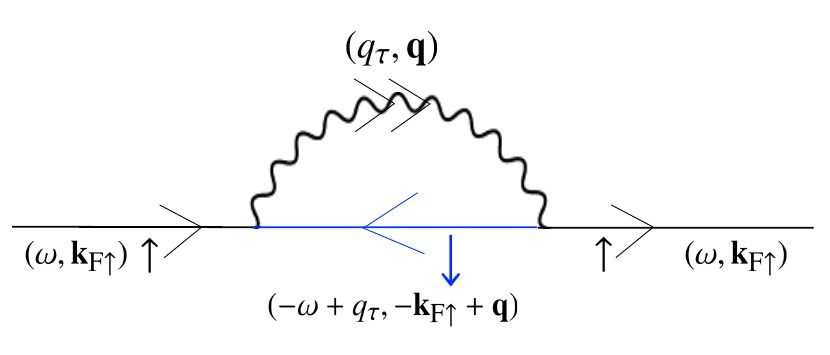}
\caption{
(Color online) Infrared finite one-loop 
contribution to the fermion self-energy. The internal 
(blue) fermion line is always evaluated off-shell, that is, 
away from its Fermi surface. Note that in the different LOFF-case, where the Cooperon 
carries finite momentum, already this diagram can induce non-Fermi liquid behavior.
This case is further discussed in Sec.~\ref{sec:key}.
}
\label{fig:fock}
\end{figure}

We now analyze the structure of the repeated 
scattering mechanism off finite ($q_\tau$, $\mathbf{q}$) fluctuations at the quantum 
critical point. The single-scattering event of Fig.~\ref{fig:fock} is infrared finite because the 
respective, internal fermion line is always evaluated off-shell, 
that is, away from its Fermi surface. Replacing the blue (spin-down) fermion line 
with a constant, the phase measure from the $q_\tau$ and $\mathbf{q}$ 
integration yields an (infra-red) finite integral over the massless Cooper 
pair propagator with $z=2$ dynamics.
The 
leading infrared-dangerous scattering process for the 
frequency-dependent self-energy of the spin-up Fermi surfaces,
$\Sigma_{\uparrow}\left(\omega,\mathbf{k}_{\rm F\uparrow}\right)$, 
is a two-loop contraction shown in the top of Fig.~\ref{fig:diagrams}:
\begin{align}
\Sigma_{\uparrow}\left(\omega,\mathbf{k}_{\rm F \uparrow}\right)
&=
g^4 \int_{\boldsymbol{\ell},\ell_\tau} 
\Bigg\{
 \int_{\boldsymbol{q},q_\tau} 
 \left[G_{\downarrow}(-\omega+q_\tau,-\mathbf{k}_{\text{F}\uparrow}+\mathbf{q})\right]^2
\nonumber\\
&
\hspace{25mm}D(q_\tau,\mathbf{q})
D(q_\tau+\ell_\tau,\mathbf{q}+\boldsymbol{\ell})
\Bigg\}\times
\nonumber\\
&\hspace{15mm}G_{\uparrow}(\omega+\ell_\tau,\mathbf{k}_{\text{F}\uparrow}+\boldsymbol{\ell})
\nonumber\\
&\equiv
g^4 \int_{\boldsymbol{\ell},\ell_\tau} 
\Bigg\{
C(\ell_\tau, \boldsymbol{\ell})
\Bigg\}
G_{\uparrow}(\omega+\ell_\tau,\mathbf{k}_{\text{F}\uparrow}+\boldsymbol{\ell})\;,
\end{align}
where we will view the function $C(\ell_\tau, \boldsymbol{\ell})$ as a strongly frequency- and 
momentum dependent {\it intra-species} interaction. 
In local coordinate 
systems around the Fermi surfaces (cf. Fig.~\ref{fig:spheres}), the (on-shell) fermion propagator can be expressed as
\begin{align}
G_{\uparrow}(\omega+ \ell_\tau,\mathbf{k}_{\rm F \uparrow} + \boldsymbol{\ell})
=\frac{1}{-i (\omega + \ell_\tau) + \rm{v}_{\rm F \uparrow} \ell_{\perp}
+\frac{1}{2 m_{\uparrow}} \left(\ell_{\perp}^2 + \ell_{\parallel}^2\right)}
\label{eq:fermi_prop}
\end{align}
with Fermi velocity $\rm v_{\rm F \uparrow} = \sqrt{\frac{\mu_\uparrow}{2m_\uparrow}}$.
The $\ell_{\perp}^2$ term is subdominant in the infrared compared to the linear 
$\rm{v}_{\rm F\uparrow}\ell_\perp$ term.
The Cooper pair propagator at criticality is
\begin{align}
D(q_\tau,\mathbf{q})=\frac{1}{-iq_\tau + q_\tau^2 + \mathbf{q}^2}
\label{eq:D}
\end{align}
where we may set the $X$- and $Z$-factors to unity here. 
Note that in an effectively particle-hole symmetric situations 
(in symmetric shells around the Fermi surfaces), the complex 
linear frequency term must vanish and the bare dynamical exponent would be $z=1$.
With disorder from trapping inhomogeneities or an additional species 
off which the Cooper pairs could scatter, one would have overdamped dynamics 
$\sim | q_\tau|$ \cite{delmaestro09}.

Evaluating 
the {\it spin-down} fermions (the blue lines in the red box of Fig.~\ref{fig:diagrams})
at the {\it spin-up} Fermi surface $\mathbf{k}_{{\rm F}\uparrow}$ makes them 
off-shell and 
the infrared-dominant contribution of the dynamical intra-species interaction 
then comes from the two Cooper pair propagators 
\begin{align}
C_{\text{IR}}(\ell_\tau, \boldsymbol{\ell})
&\sim 
\int \frac{d^2 \mathbf{q}}{(2 \pi)^2} \int \frac{d q_\tau}{2\pi}
D(q_\tau,\mathbf{q})
D(q_\tau+\ell_\tau,\mathbf{q}+\boldsymbol{\ell})
\label{eq:DD}
\end{align}
which, for example, is logarithmically singular with a bare $z=2$ power-counting.
The precise form of the effective single-species interaction mediated by density fluctuations and 
two critical Cooperons also needs to account for the non-analyticities from pair-breaking effects from 
logarithm of the Cooperon, Eq.~(\ref{eq:log}), when summing over higher energy intermediate states 
as well as more precise experimental considerations such as trapping potentials and the presence of disorder. 
Upon closing this singular interaction with the on-shell spin-up fermion propagator Eq.~(\ref{eq:fermi_prop}) 
we conclude that the lifetime of quasiparticles cannot have the Fermi liquid form (where the interactions 
are regular and IR finite and $  \text{Im}\Sigma_{\uparrow}\left(\omega,\mathbf{k}_{\rm F \uparrow}\right) 
\sim \omega^2$) but instead will receive anomalous corrections.
%
%

A similar mechanism was shown to induce non-Fermi behavior 
in antiferromagnetic metals recently \cite{hartnoll11}. 
Strong non-Fermi liquid behavior (``hot Fermi surfaces") and the total destruction of quasiparticles occurs for 
would require a strong-coupling re-summation of the 
logarithmic singularities of Eq.~(\ref{eq:DD}) as found by Abrahams {\it et al.} \cite{abrahams13} in the 
spin-fermion model; such a re-summation would effectively lead to power-law, long-ranged interaction 
\`a la Bares and Wen \cite{bares93}.

Compared with the spin-density wave and Ising nematic quantum critical points in Table \ref{fig:table}, 
the homogeneous imbalanced superfluid discussed in the present paper is expected to feature 
warm Fermi surfaces leading to anomalous thermodynamics and transport but, in the absence of 
band crossings/intersections of the two Fermi surfaces, ultimately stable quasi-particles.
At the hot spots in the spin-density wave case ($\text{Im}\Sigma\sim\sqrt{\omega}$), 
or in the anti-nodal region of the nematic case ($\text{Im}\Sigma\sim\omega^{2/3}$), one 
has truly hot regions without any quasiparticles at all.

\section{Conclusion and outlook on Experiments }
\label{sec:conclusion}

This paper has two main messages. First, we showed that quantum fluctuations 
may actually favor the appearance of exotic superfluids in two-dimensional, 
imbalanced mixtures of fermionic atoms. Second, our results put imbalanced 
fermionic superfluids on the map of plain vanilla quantum phase transitions (Fig.~\ref{fig:table}) 
in two spatial dimensions exhibiting non-Fermi liquid features. In light of the experimental progress 
in homogeneous trapping of fermionic atoms in reduced dimensionality, this paper  
hopefully stimulates that the results of mean-field approaches (applicable mostly to three-dimensional 
gases), will be carefully revisited in experiments and theory in two spatial dimensions.

Order parameter fluctuations, 
captured by solving a partial differential flow equation for the effective 
potential of the imbalanced Fermi gas, 
were shown to have dramatic beyond-mean-field effects on the zero-temperature phase diagram: 
for a range of parameters, a new, fluctuation-induced quantum critical point 
was found and Goldstone fluctuations were shown to destroy spontaneous 
symmetry breaking completely, in accordance with the Mermin-Wagner theorem.

We argued that this case is qualitatively 
different from the more frequently studied case of fermions coupled to gauge field and the 
hot spot phenomenology at the onset of antiferromagnetic order in metals. Its key features are:

(i) The collective mode(s) appears in the pairing channel spontaneously breaking 
the global $U(1)$ symmetry. This leads to a single Goldstone mode whose 
fluctuations were shown to turn the mean-field phase-separated ground state into a homogeneous, 
strongly coupled imbalanced superfluid with gapless fermion excitations. This also leads to different features 
 in the quantum-critical field theory 
coupling two mismatched Fermi surfaces to the Cooperon. This bosonic Cooperon, for example, has a complex, linear time-derivative 
that leads to a non-trivial coupling of Goldstone and longitudinal couplings in the superfluid phase. It may additionally be 
Landau-damping from small disorder or trap inhomogeneities. Moreover, 
at the quantum critical point, one-loop vertex corrections renormalizing the Yukawa-coupling vanish by 
particle conservation - these structural novelties make a comprehensive, diagrammatic analysis 
of the coupled Fermi-Bose theory an interesting topic for future research. 

(ii) Non-Fermi liquid physics around the Fermi surfaces appears although the Cooperon 
does not directly connect points or patches in the two mismatched Fermi surfaces. Repeated 
scattering processes should lead to warm Fermi surfaces with reduced, non-Fermi liquid quasi-particle lifetimes. 

(iii) The microscopic model of the parent compound is directly realizable in present-day experiments with 
ultracold mixtures of fermionic atoms. Techniques such as rf-spectroscopy \cite{stewart08,lan13} bear the potential 
to measure the reduced fermion lifetimes. The frequency resolved rf-signal is proportional to the overlap 
of a non-interacting fermion propagator from a non-interacting third level with the interacting 
fermion and therefore should be able to resolve the single-particle self-energy \cite{haussmann09}. Furthermore, 
one can measure the frequencies of collective Rabi oscillations directly which can be related to the quasi-particle weight, too 
\cite{grimm12}.

The QCP's of Fig.~\ref{fig:domes} should also have observable consequences such as anomalous 
thermodynamics \cite{loehneysen07} up to relatively, high, finite 
temperatures in the quantum critical regime, even in a scenario in which it gets ``masked'' by p-wave superfluidity 
at the lowest scales. Finally, the collective modes such as sound and the longitudinal ``Higgs'' mode can also be measured 
for example by time-modulating the trapping potential and measuring the associated density response of the gas \cite{koehl12}. 
It would therefore be of great interest to generalize 
recent experiments with two-dimensional Fermi gases \cite{koehl11a,koehl11b} 
to Fermi mixtures \cite{grimm12} in the superfluid regime.

\begin{acknowledgments}
We acknowledge helpful discussions with Bertrand Delamotte, Nicolas Dupuis, Matthias Punk, 
Subir Sachdev, Brian Swingle, Tarik Yefsah, and Martin Zwierlein.
The research of PS was supported by the DFG under grant Str 1176/1-1, by the NSF under Grant DMR-1103860, 
by the John Templeton Foundation, by the Center for Ultracold Atoms (CUA) and by 
the Multidisciplinary University Research Initiative (MURI). PJ acknowledges funding by the National Science Centre via 2011/03/B/ST3/02638.
\end{acknowledgments}

\begin{appendix}

\section{Linking the flow to fermionic correlations}
\label{app:mft}

We here present the derivation of the mean-field effective potential Eq.~(\ref{eq:local_pot}) 
and the order parameter propagators  Eq.~(\ref{eq:matrix_element_props}) thereby 
providing a direct link of the order parameter flow to the underlying fermionic atoms.

We go to the thermodynamic limit and employ an action representation for Eq.~(\ref{eq:hamiltonian}). 
We perform a Hubbard-Stratonovich transformation with the spin-singlet pairing 
field $\phi_q$ conjugate to the fermion bilinear
\begin{equation}
 \tilde\phi_{q} = g \int_k \psi_{k+\frac{q}{2}\uparrow}
 \psi_{-k+\frac{q}{2}\downarrow} \; .
\end{equation}
We employ the short-hand notation $\int_k=T\sum_{k_0}\int\frac{d^2\bold{k}}{(2\pi)^2}$ with fermionic 
(imaginary) Matsubara frequencies $k_0=2\pi (n+1/2)T$ with $n$ integer running from $-\infty$ to $\infty$. 
Their bosonic pendant is conventionally denoted by $q_0=2\pi n T$. Later we will take the zero temperature 
limit $T\rightarrow 0$ limit and focus on ground state properties.
The resulting functional integral is over $\psi$, $\bar{\psi}$ and $\phi$
with the new bare action
\begin{eqnarray}
 \mathcal{S}[\psi,\bar{\psi},\phi]
  &=& \int_{k}\sum_{\sigma=\uparrow,\downarrow} \bar{\psi}_{k\sigma} (-ik_{0} + \xi_{\bold{k}\sigma}) \, \psi_{k\sigma}
  - \int_q \phi^*_q \frac{1}{g} \phi_q \nonumber \\
  && + \int_{k,q} \left( 
  \bar{\psi}_{-k+\frac{q}{2} \downarrow}\bar{\psi}_{k+\frac{q}{2} \uparrow} \,
  \phi_{q} +
  \psi_{k+\frac{q}{2} \uparrow} \psi_{-k+\frac{q}{2}\downarrow} \,
  \phi^*_q \right),
\nonumber\\\label{eq:action}
\end{eqnarray}
where $\phi^*$ is the complex conjugate of $\phi$, while $\psi$
and $\bar{\psi}$ are algebraically independent Grassmann variables. Note that 
as $g<0$, the bosonic mass term -1/g is positive. 

We first shift $\phi_q$ by a non-fluctuating (static in time and homogenous in position space) field $\Delta$:
\begin{align}
\phi_q&\rightarrow(2\pi)^2\delta_{\bold{q},0}\frac{\delta_{\Omega,0}}{T}\,\Delta+\phi_q\nonumber\\
\label{eq:phi_0}
\end{align}
where the factors of $(2\pi)^2$ are conventional and 
translate to the correct volume normalization factors in position space. 
Eq.~(\ref{eq:action}) is recast as $\mathcal{S}=\mathcal{S}_{b}+\mathcal{S}_f$ with
\begin{align}
\mathcal{S}_{b}&=\frac{(2\pi)^2}{T}|\Delta|^2\left(\frac{-1}{g}\right)
+
\left(\frac{-1}{g}\right)\left(\Delta\phi_0^\ast+\Delta^\ast\phi_0\right)
+\int_q\phi_q^\ast\phi_q\left(\frac{-1}{g}\right)\;.
\label{eq:S_b}
\end{align}
We employ a convenient Nambu matrix notation for the fermionic part
\begin{align}
\mathcal{S}_{f}
&=
\int_k \left(\bar{\psi}_{k\uparrow}\;\psi_{-k\downarrow}\right)
\overbrace{
\left( \begin{array}{cc}
-i k_0 + \xi_{\bold{k}\uparrow}  & -\Delta \\
-\Delta^\ast & -ik_0 - \xi_{-\bold{k}\downarrow} 
\end{array} \right)
}^{\equiv\mathcal{S}^{(2)}_f\left[\Delta\right]}
\left( \begin{array}{c}
\psi_{k\uparrow}\\
\bar{\psi}_{-k\uparrow}
\end{array} \right)
\nonumber\\
&+
\int_{k,q} \left(\bar{\psi}_{k+\frac{q}{2}\uparrow}\;\psi_{-k+\frac{q}{2}\downarrow}\right)
\left( \begin{array}{cc}
0 & -\phi_q \\
-\phi_q^\ast & 0
\end{array} \right)
\left( \begin{array}{c}
\psi_{k+\frac{q}{2}\uparrow}\\
\bar{\psi}_{-k+\frac{q}{2}\uparrow}
\end{array} \right)\;.
%
%
\label{eq:nambu_action}
\end{align}
This action is quadratic in the Grassmann fermion fields and we can perform the 
Gaussian functional integral. 
After expanding the resulting tr $\log$ to second order in $\phi_q$, we obtain
\begin{align}
\mathcal{S}[\phi]=&\int_q\phi_q^\ast\left(\frac{-1}{g}+K(q)\right)\phi_q+
\int_q\phi^\ast_q\phi^\ast_{-q}\frac{L(q)}{2}
+\int_q\phi_q\phi_{-q}\frac{L^\ast(q)}{2}
\nonumber\\
&-\frac{(2\pi)^2}{T}\frac{1}{g}|\Delta|^2-\text{tr}\ln\beta \mathcal{S}^{(2)}_{f}[\Delta]\nonumber\\
& + \left(\frac{-1}{g}\right)
\left(\Delta\phi_0^\ast+\Delta^\ast\phi_0\right)
-\left(\phi_0^\ast\int_k F_{f\uparrow\downarrow}(k)
+\phi_0\int_k F^\ast_{f\uparrow\downarrow}(k)
\right)\;,
\label{eq:phi_action}
\end{align}
where the tr goes over Nambu indices, frequencies and momenta.
The terms in the last line cancel, if $\Delta$ fulfills the saddle-point condition or gap equation
\begin{equation}
\Delta=-g\int_k F_{f\uparrow\downarrow}(k)\;,
\end{equation}
which is what one gets from differentiating the second line of Eq.~(\ref{eq:phi_action}) with respect to $\Delta^\ast$.
Here, we have used the normal ($K$) and anomalous ($L$) particle-particle pair propagators
\begin{align}
K(q)=-\int_k G_{f\uparrow}(k+\frac{q}{2})G_{f\downarrow}(-k+\frac{q}{2})\nonumber\\
L(q)=\int_k F_{f\uparrow\downarrow}(k+\frac{q}{2})F_{f\downarrow\uparrow}(-k+\frac{q}{2})\;,
\label{eq:pair_props}
\end{align}
and their single-particle constituents
\begin{align}
G_{f\uparrow}(k)&=\frac{-i k_0 - \xi_{-\bold{k}\downarrow}}
{(ik_0-\xi_{\bf{k}\uparrow})(-ik_0-\xi_{-\bf{k}\downarrow})+|\Delta|^2}\nonumber\\
F_{f\uparrow\downarrow}(k)&=\frac{\Delta}
{(ik_0-\xi_{\bf{k}\uparrow})(-ik_0-\xi_{-\bf{k}\downarrow})+|\Delta|^2}\;.
\label{eq:single_props}
\end{align}

To bring out the physics of Goldstone and longitudinal fluctuations, 
we now perform a change of variables in Eq.~(\ref{eq:phi_action}), $\mathcal{S}[\phi]\rightarrow\mathcal{S}[\sigma,\pi]$, 
with
\begin{align}
\phi_q&=\frac{\sigma_q+i\pi_q}{\sqrt{2}}\nonumber\\
\phi^\ast_q&=\frac{\sigma_{-q}-i\pi_{-q}}{\sqrt{2}}\;,
\end{align}
where $\pi$ is the transversal Goldstone field variable and $\sigma$ the radial Higgs field variable. 
$\Delta$ is chosen real and denoted by $\alpha$ such that 
\begin{align}
|\Delta|^2=\frac{\alpha^2}{2}\equiv\rho\;.
\label{eq:def_rho}
\end{align}
The action obtains as
\begin{align}
\mathcal{S}[\sigma,\pi,\alpha]&=\int_q \frac{\sigma_{-q}\sigma_q}{2}\left[-g^{-1}+K(q)+L(q)\right]
\nonumber\\
&+
\int_q \frac{\pi_{-q}\pi_{q}}{2}\left[-g^{-1}+K(q)-L(q)\right]\nonumber\\
&+
\int_q\frac{\sigma_{-q}\pi_q}{2}
\left(i K^{\text{odd}}(q)+i \frac{L^{\ast\text{odd}}(q)-L^{\text{odd}}(q)}{2}\right)
\nonumber\\
&+ 
\frac{\pi_{-q}\sigma_{q}}{2}
\left(-i K^{\text{odd}}(-q)+i \frac{L^{\ast\text{odd}}(-q)-L^{\text{odd}}(-q)}{2}\right)
\nonumber\\
&
+\frac{(2\pi)^2}{T} U[\alpha]\;,
\label{eq:action_sigma_pi}
\end{align}
with $U[\alpha]$ the local effective potential given in Eq.~(\ref{eq:local_pot}) that 
survives $q\rightarrow 0$ mean-field limit.
The mean-field phase diagram of Fig.~\ref{fig:domes} is 
determined by the field value $\alpha$ that {\em globally} 
minimizes the potential
\begin{align}
\Omega^0(\mu,h,r,T)=\text{min}_{\alpha}
\Big\{
U[\alpha]
\Big\}\;.
\label{eq:free_energy_density}
\end{align}
The $q\neq 0$ terms composed of fermionic pair propagators represent phase 
and amplitude fluctuations 
about the mean-field saddle-point. The superscript `odd' projects out frequency and momentum 
components that transform odd under $q\rightarrow -q$. Because the anomalous pair propagator 
$L(q)$ of Eq.~(\ref{eq:pair_props}) does not have an imaginary part, that is, $L^\ast(q)=L(q)$, the terms $\sim L$ in Eq.~(\ref{eq:action_sigma_pi}) cancel.

The mass terms of the radial and Goldstone propagators can be obtained from derivatives 
of $U[\alpha]$. For the Goldstone propagator, one gets
\begin{align}
\frac{1}{\alpha}U'[\alpha]\equiv\frac{1}{\alpha}\frac{\partial U}{\partial \alpha}&=\frac{-1}{g}-
\frac{1}{\alpha }\int_k F_{f\uparrow\downarrow}(k)
=\frac{-1}{g}+K(0)-L(0)\;,
\label{eq:m_pi}
\end{align}
where the second line uses the explicit expression of the pair propagators in Eq.~(\ref{eq:pair_props}). 
The Goldstone mass is zero if $U'[\alpha_0]=0$ or, equivalently, $\alpha_0$ fulfills the saddle-point condition or 
gap equation.
For the radial propagator, one gets
\begin{align}
U''[\alpha]=\frac{\partial^{2}U}{\partial \alpha^2}&=\frac{1}{\alpha}\frac{\partial U}{\partial \alpha}+ 2 L(0)
=\frac{-1}{g}+K(0)+L(0)\;,
\label{eq:m_sigma}
\end{align}
where we use Eq.~(\ref{eq:m_pi}) to obtain the second line. $U''[\alpha_0]$ is finite providing the 
mass gap to the longitudinal fluctuations. We finally rewrite Eq.~(\ref{eq:action_sigma_pi}) using 
Eqs.~(\ref{eq:m_pi},\ref{eq:m_sigma}) to obtain: 
\begin{align}
\mathcal{S}[\sigma,\pi,\alpha]=&\int_q \frac{\sigma_{-q}\sigma_q}{2}\left(Q_{\sigma\sigma}(q;\alpha) + U''[\alpha]\right)
+\frac{(2\pi)^2}{T} U[\alpha]\nonumber\\
&+
\int_q \frac{\pi_{-q}\pi_{q}}{2}\left(Q_{\pi\pi}(q;\alpha)+\frac{1}{\alpha}U'[\alpha]\right)
\label{eq:prefinal_action}\\
&
+\int_q \left(\frac{\sigma_{-q}\pi_q}{2} Q_{\sigma\pi}(q;\alpha) + \frac{\pi_{-q}\sigma_q}{2} Q_{\sigma\pi}(-q;\alpha) \right)\;,
\nonumber
\end{align}
with the momentum behavior of the propagators 
\begin{align}
Q_{\sigma\sigma}(q)&=K(q)+L(q)-(K(0)+L(0))\nonumber\\
Q_{\pi\pi}(q)&=K(q)-L(q)-(K(0)-L(0))\nonumber\\
Q_{\sigma\pi}(q)&=i K^{\text{odd}}(q)\;.
\label{eq:Q_functions}
\end{align}
Clearly, $Q_{\sigma\sigma}(q=0)=Q_{\pi\pi}(q=0)= Q_{\sigma\pi}(q=0)=0$.

Eqs.~(\ref{eq:Q_functions}) tie the renormalization group analysis for order parameter 
fluctuations Eq.~(\ref{eq:matrix_element_props}) to the underlying fermionic correlations.

\section{Details of the RG flow}
\label{app:details}

We here give some details about the applied RG procedure.
We restrict the momentum 
integrations to $|\mathbf{k}| < 3.5 |\mathbf{k}_{\rm F}| = 3.5 \times \sqrt{2}$ mimicking the UV-regularization 
of an attractive lattice Hubbard model at low filling fractions. In our computations, the initial momentum scale 
$\Lambda_0$ at which we start the flow is a free parameter. Generally, this should be momentum scale below which 
order parameter fluctuations are expected to kick in and we here take 
$\Lambda_0 = \sqrt{2} \sim |\mathbf{k}_{\text F}|$ attempting to account for an entire class of 
particle-hole and higher-order diagrams which are formally omitted within our truncation. These are expected to reduced ordering scales significantly \cite{gorkov61}.

The applied procedure is an adapted variant of a leading-order calculation in the functional  renormalization-group 
derivative expansion. The next-to-leading order would also treat the $Z$- and 
$X$-factors as functions of $\rho$ and $\Lambda$. Our main focus here are the zero-temperature flows leading to the 
phase diagram in Fig.~\ref{fig:domes}.  For this situation a reliable approximation
\footnote{
On the other hand, if taken into account, anomalous scaling of the propagator occurs exclusively at the 
transition and only once the fixed point is attained (at low $\Lambda$). 
Therefore our approximation should have no impact on our conclusions.
}
neglects the order parameter flow of 
$Z$ and $X$ altogether and leaves their initial values at $\Lambda=\Lambda_0$ unchanged from the values guided by the form 
of the fermionic theory, Eqs.~(\ref{eq:Q_functions}), as well as attempting to account for 
particle-hole fluctuations and higher-order corrections. We here chose
$Z^{\Lambda_0}_{\Omega}=Z^{\Lambda_0}_{\mathbf{q}}=1$, and $X^{\Lambda_0}=0.2$. 

\section{Imbalanced particle-particle bubble}
\label{app:ppbubble}

The imbalanced particle-particle bubble drawn in Fig.~\ref{fig:ppbubble}, whose low-energy expansion leads to 
the Cooper pair propagator Eqs.~(\ref{eq:phiphi},\ref{eq:D}), is defined as
\begin{equation}
\Pi(q)=\Pi_{\text{pp}}(iq_\tau,\mathbf{q})-\Pi_{\text{pp}}(0,0)\;,
\label{eq:bosonic_prop}
\end{equation}
with 
\begin{equation}
\Pi_{\text{pp}}(iq_\tau,\mathbf{q})
=-
\int^{\infty}_{-\infty}\frac{d k_\tau}{2\pi}
\int^{\infty}_{-\infty}\frac{d^2 \mathbf{k}}{(2\pi)^2}
G_{\uparrow}(k+q/2)G_{\downarrow}(-k+q/2)\;.
\label{eq:pp_bubble}
\end{equation}
and
\begin{equation}
G_{\sigma}(k)=\frac{1}{-i k_\tau +  \xi_{\mathbf{k}\sigma}}
=\frac{1}{-i k_\tau + \frac{\mathbf{k}^2}{2m_{\sigma}}-\mu_{\sigma}}
\end{equation}
We can successively perform the 
integrations first $k_\tau$, then $k_x$, and $k_y$ last to find
\begin{align}
\Pi\left(q\right)
=
\frac{-1}{4\pi} \log\left[1+\frac{Q(q)}{2\mu}\right]\;.
\label{eq:log}
\end{align}
where $\mu=\frac{1}{2}\left(\mu_{\uparrow} + \mu_{\downarrow}\right)$
and
\begin{align}
Q(q)=Q(q_\tau,\mathbf{q}) = i q_\tau- \mathbf{q}^2\left( \frac{1}{4 m_r}-\frac{m_r}{16}
\left(\frac{1}{m_\uparrow} - \frac{1}{m_\downarrow}\right)^2\right)\;.
\label{eq:Q}
\end{align}
The logarithm in Eq.~(\ref{eq:log}) will in general lead to branch cuts in the complex (frequency plane). 
For energies and momenta smaller than the effective binding energy $2\mu$, we can expand 
the logarithm to obtain Eq.~(\ref{eq:phiphi}) with
\begin{align}
X&=\frac{1}{8\pi\mu}\nonumber\\\
Z_{\Omega} &= \frac{1}{32\pi\mu^2}\\
Z_{\mathbf{q}}&=\frac{1}{32\pi\mu m_r}\left(1-\frac{m_r^2}{4}\left(\frac{1}{m_{\uparrow}}-\frac{1}{m_{\downarrow}}\right)^2\right)\;,
\label{eq:Zs_normal}
\end{align}
where we further used the reduced mass $\frac{1}{m_r}=\frac{1}{2}\left(\frac{1}{m_{\uparrow}}+\frac{1}{m_{\downarrow}}\right)$.

\end{appendix}


\newpage
\begin{thebibliography}{99}

\bibitem{landau56}
L. D. Landau, 
{\it Theory of Fermi-liquids}, Sov. Phys. JETP {\bf 3}, 920 (1957);
{\it Oscillations in a Fermi-liquid}, ibid {\bf 5}, 101 (1957).

\bibitem{nozieres64} 
P. Nozi\`eres, {\it Theory of Interacting Fermi Systems}, 
Benjamin, Amsterdam (1964).

\bibitem{schofield99}
A. J. Schofield, {\it Non-Fermi liquids},
Contemporary Physics {\bf 40}, 95 (1999).

\bibitem{varma02}
C. M. Varma, Z. Nussinov, W. van Saarloos,
{\it Singular or non-Fermi liquids}, 
Phys. Repts {\bf 361}, 267 (2002).

\bibitem{schaefer04}
T. Schaefer, and K. Schwenzer,
{\it Non-Fermi liquid effects in QCD at 
high density}, 
Phys. Rev. D {\bf 70}, 054007 (2004).

\bibitem{loehneysen07}
H. v. Loehneysen, A. Rosch, M. Vojta, 
and P. Woelfle, 
{\it Fermi-liquid instabilities at magnetic quantum phase transitions}, 
Rev. Mod. Phys. {\bf 79}, 1015 (2007).

\bibitem{schofield05}
P. Coleman, and A. J. Schofield,
{\it Quantum criticality},
Nature {\bf 433}, 226 (2005).

\bibitem{sachdev11}
S. Sachdev, and B. Keimer, 
{\it Quantum criticality}, 
Physics Today {\bf 64}, 29 (2011).

\bibitem{greywall83}
D. S. Greywall, 
{\it Specific heat of normal liquid $^{3}$ He}, 
Phys. Rev. B {\bf 27}, 2747 (1983).

\bibitem{vollhardt03}
D. Vollhardt, P. Woelfle,
{\it The Superfluid Phases Of Helium 3}, 
Taylor \& Francis (2003).

\bibitem{zwerger08}
I. Bloch, J. Dalibard, and W. Zwerger,
{\it Many-body physics with ultracold gases},
Rev. Mod. Phys. {\bf 80}, 885 (2008).

\bibitem{jin99}
B. DeMarco, and D. S. Jin, 
{\it Onset of Fermi Degeneracy in a Trapped 
Atomic Gas}, Science {\bf 285}, 1703 (1999).

\bibitem{zwierlein03}
Z. Hadzibabic {\it et al.}, 
{\it Fiftyfold Improvement in the Number of Quantum 
Degenerate Fermionic Atoms}, 
Phys. Rev. Lett. {\bf 91}, 160401 (2003).

\bibitem{koehl05}
M. Koehl {\it et al.}, 
{\it Fermionic atoms in three dimensional optical lattice:
Observing Fermi surfaces, dynamics, and interactions}, 
Phys. Rev. Lett. {\bf 94}, 80403 (2005).

\bibitem{zwierlein06}
M. Zwierlein {\it et al.}, 
{\it Fermionic Superfluidity with Imbalanced Spin Populations}, 
Science {\bf 311}, 492 (2006).

\bibitem{partridge06}
G. B. Partridge {\it et al.}, 
{\it Pairing and Phase Separation in a Polarized Fermi Gas}, 
Science {\bf 311}, 501 (2006).

\bibitem{grimm12}
C. Kohstall {\it et al.}, 
{\it Metastability and coherence of repulsive 
polarons in a strongly interacting Fermi mixture},
Nature {\bf 485}, 615 (2012).

\bibitem{liu03}
W. V. Liu, and F. Wilczek, 
{\it Interior Gap Superfluidity},
Phys. Rev. Lett. {\bf 90}, 047002 (2003).

\bibitem{sarma63}
G. Sarma,
{\it On the influence of a uniform exchange field 
acting on the spins of the conduction electrons 
in a superconductor},
J. Phys. Chem. Solids, {\bf 24}, 1029 (1963).

\bibitem{bedaque03}
P. F. Bedaque, H. Caldas, and G. Rupak,
{\it Phase Separation in Asymmetrical 
Fermion Superfluids}, 
Phys. Rev. Lett. {\bf 91}, 247002 (2003).

\bibitem{yip03} S.-T. Wu and S. Yip,
{\it Superfluidity in the interior-gap states},
Phys. Rev. A {\bf 67}, 053603 (2003).

\bibitem{liu04}
W. V. Liu, F. Wilczek, and P. Zoller,
{\it Spin-dependent Hubbard model and a quantum 
phase transition in cold atoms},
Phys. Rev. A {\bf 70}, 033603 (2004).

\bibitem{schwenk06}
A. Bulgac, M. McNeil Forbes, and A. Schwenk,
{\it Induced P-Wave Superfluidity in Asymmmetric Fermi Gases},
Phys. Rev. Lett. {\bf 97}, 020402 (2006).

\bibitem{parish07} M. M. Parish, F. M. Marchetti, A. Lamacraft, and 
B. D. Simons, 
{\it Polarized Fermi Condensates with Unequal Masses: Tuning the 
Tricritical Point}, Phys. Rev. Lett. {\bf 98}, 160402 (2007).

\bibitem{he08}
L. He, and P. Zhuang,
{\it Phase diagram of a cold polarized Fermi gas in two dimensions},
Phys. Rev. A {\bf 78}, 033613 (2008).

\bibitem{pilati08}
S. Pilati, and S. Giorgini,
{\it Phase Separation in a Polarized Fermi Gas at Zero Temperature},
Phys. Rev. Lett. {\bf 100}, 030401 (2008).

\bibitem{fisher09}
A. E. Feiguin, and M. P. A. Fisher,
{\it Exotic Paired States with Anisotropic Spin-Dependent 
Fermi Surfaces},
Phys. Rev. Lett. {\bf 103}, 025303 (2009).

\bibitem{randeria10}
R. B. Diener, and M. Randeria,
{\it BCS-BEC crossover with unequal-mass fermions},
Phys. Rev. A {\bf 81}, 033608 (2010).

\bibitem{sheehy10}
L. Radzihovsky, and D. E. Sheehy,
{\it Imbalanced Feshbach-resonant Fermi gases},
Rep. Prog. Phys. {\bf 73}, 076501 (2010).

\bibitem{cichy11} 
A. Kujawa-Cichy and R. Micnas 
{\it Stability of superfluid phases in the 2D Spin-Polarized Attractive Hubbard Model}
Europhys. Lett. {\bf 95}, 37003 (2011).

\bibitem{sheehy09}
D. E. Sheehy,
{\it Polarized superfluids near their tricritical point},
Phys. Rev. A {\bf 79}, 033606 (2009)

\bibitem{torma13}
S. Yin, J.-P. Martikainen, and P. Torma,
{\it Fulde-Ferrel states and Berezinskii-Kosterlitz-Thouless 
phase transition in two-dimensional imbalanced Fermi gases},
Phys. Rev. B {\bf 89}, 014507 (2014).

\bibitem{clogston62} A. M. Clogston,
{\it Upper limit for the critical field in hard superconductors},
Phys. Rev. Lett. {\bf 9}, 266 (1962).

\bibitem{Binder87} K. Binder 
{\it Theory of first-order phase transitions}, 
Rep. Prog. Phys. {\bf 50}, 783 (1987).
  
\bibitem{Lohneysen07} H. v. L\"{o}hneysen, A. Rosch, M. Vojta, and P. W\"{o}lfle, 
{\it Fermi-liquid instabilities at magnetic quantum phase transitions},
Rev. Mod. Phys. {\bf 79}, 1015 (2007). 

\bibitem{Belitz05} D. Belitz, T. R. Kirkpatrick, and T. Vojta, 
{\it How generic scale invariance influences quantum and classical phase transitions},
Rev. Mod. Phys. {\bf 77}, 579 (2005).  

\bibitem{Greenblatt09} R. L. Greenblatt, M. Aizenman, and J.L. Lebowitz, 
{\it Rounding of First Order Transitions in Low-Dimensional Quantum Systems with Quenched Disorder}, 
Phys. Rev. Lett. {\bf 103}, 197201 (2009). 

\bibitem{Jakubczyk10} P. Jakubczyk, J. Bauer, and W. Metzner
{\it Finite temperature crossovers near quantum tricritical points in metals}, 
Phys. Rev. B {\bf 82}, 045103 (2010). 
 
\bibitem{stewart08}
J. T. Stewart, J. P. Gaebler, and D. S. Jin,
{\it Using photoemission spectroscopy to probe 
a strongly interacting Fermi gas},
Nature {\bf 454}, 744 (2008).

\bibitem{hulet10}
Y. Liao {\it et al.}, 
{\it Spin-imbalance in a one-dimensional Fermi gas},
Nature {\bf 467}, 567 (2010).

\bibitem{lamacraft10}
A. J. A. James, and A. Lamacraft,
{\it Non-Fermi-Liquid Fixed Point for an Imbalanced Gas of 
Fermions in 1+$\epsilon$ Dimensions},
Phys. Rev. Lett. {\bf 104}, 190403 (2010).

\bibitem{sheehy07}
D. E. Sheehy, and L. Radzihovsky,
{\it BEC-BCS crossover, phase transitions and 
phase separation in polarized resonantly-paired superfluids},
Annals of Physics {\bf 322}, 1790 (2007).

\bibitem{lutchyn11}
R. M. Lutchyn, M. Dzero, and V. M. Yakovenko,
{\it Spectroscopy of the soliton lattice formation in 
quasi-one-dimensional fermionic superfluids with 
population imbalance}, 
Phys. Rev. A {\bf 84}, 033609 (2011).


\bibitem{scalettar12}
M. J. Wolak, B. Gremaud, R. T. Scalettar, and 
G. G. Batrouni,
{\it Pairing in two-dimensional Fermi gas with population imbalance}, 
Phys. Rev. A {\bf 86}, 023630 (2012).

\bibitem{roscher13}
D. Roscher, J. Braun, and J. E. Drut,
{\it Inhomogeneous phases in one-dimensional mass- and 
spin-imbalanced Fermi gases},
arXiv:1311.0179 (2013).

\bibitem{lan13}
Z. Lan, G. M. Bruun, and C. Lobo,
{\it Quasiparticle Lifetime in Ultracold Fermionic Mixtures 
with Density and Mass Imbalance},
Phys. Rev. Lett. {\bf 111}, 145301 (2013).


\bibitem{max10}
M. A. Metlitski, and S. Sachdev, 
{\it Quantum phase transitions of metals in two spatial dimensions:
II. Spin density wave order},
Phys. Rev. B {\bf 82}, 075128 (2010).

\bibitem{jun12}
J. Lee, P. Strack, and S. Sachdev,
{\it Quantum criticality of reconstructing Fermi surfaces in 
antiferromagnetic metals}, 
Phys. Rev. B {\bf 87}, 045104 (2013).

\bibitem{abrahams13}
E. Abrahams, J. Schmalian, and P. Woelfle,
{\it Strong coupling theory of heavy fermion criticality},
arXiv:1303.3926.

\bibitem{kivelson01}
V. Oganesyan, S. A. Kivelson, and E. Fradkin, 
{\it Quantum theory of nematic Fermi fluid}, 
Phys. Rev. B {\bf 64}, 195109 (2001).

\bibitem{metzner03}
W. Metzner, D. Rohe, and S. Andergassen,
{\it Soft Fermi surfaces and Breakdown of Fermi-Liquid 
Behavior},
Phys. Rev. Lett. {\bf 91}, 066402 (2003).

\bibitem{lee09}
S.-S. Lee,
{\it Low-energy effective theory of Fermi surface 
coupled with U(1) gauge field in 2+1 dimensions},
Phys. Rev. B {\bf 80}, 165102 (2009).

\bibitem{powell05}
S. Powell, S. Sachdev, and H. P. Buechler,
{\it Depletion of Bose-Einstein condensate 
in Bose-Fermi mixtures},
Phys. Rev. B {\bf 72}, 024534 (2005).

\bibitem{hartnoll11}
S. A. Hartnoll, D. M. Hofman, M. A. Metlitski, and S. Sachdev,
{\it Quantum critical response at the onset of spin-density-wave 
order in two-dimensional metals},
Phys. Rev. B {\bf 84}, 125115 (2011).


\bibitem{pawel09} P. Jakubczyk, W. Metzner, and H. Yamase,
{\it Turning a First Order Quantum Phase Transition Continuous 
by Fluctuations: General Flow Equations and Application to 
d-Wave Pomeranchuk Instability},
Phys. Rev. Lett. {\bf 103}, 220602 (2009).



\bibitem{sheehy11}
K. R. Patton, D. E. Sheehy,
{\it Induced p-wave superfluidity in strongly interacting imbalanced 
Fermi gases},
Phys. Rev. A {\bf 83}, 051607 (R) (2011).

\bibitem{langmack12} C. Langmack, M. Barth, W. Zwerger, and 
E. Braaten,
{\it Clock Shift in a Strongly Interacting Two-Dimensional Fermi Gas},
Phys. Rev. Lett. {\bf 108}, 060402 (2012).

\bibitem{bertaina11} G. Bertaina, and S. Giorgini,
{\it BCS-BEC Crossover in a Two-Dimensional Fermi Gas},
Phys. Rev. Lett. {\bf 106} 110403 (2011).




\bibitem{bloom75}
P. Bloom, 
{\it Two-dimensional Fermi gas},
Phys. Rev. B {\bf 12}, 125 (1975).

\bibitem{randeria89}
M. Randeria, J.-M. Duan, and L.-Y. Shieh,
{\it Bound States, Cooper Pairing, and Bose 
Condensation in Two Dimensions}, 
Phys. Rev. Lett. {\bf 62}, 981 (1989).



\bibitem{leggett80}
A. J. Leggett, in {\it Modern Trends in the Theory of 
Condensed Matter}, 
edited by A. Pekalski and R. Przystawa (Springer, 
Berlin, 1980).

\bibitem{pistolesi04}
F. Pistolesi, C. Castellani, C. Di Castro, and G. C. Strinati,
{\it Renormalization group approach to the infrared behavior 
of a zero-temperature Bose system}, 
Phys. Rev. B {\bf 69}, 024513 (2004).

\bibitem{strack08} 
P. Strack, R. Gersch, and W. Metzner, 
{\it Renormalization group flow for fermionic superfluids at 
zero temperature}, Phys. Rev. B {\bf 78}, 014522 (2008).

\bibitem{wetterich93} C. Wetterich, 
{\it Exact evolution equation for the effective potential}, Phys. Lett. B {\bf 301}, 90 (1993). 





\bibitem{berges02} J. Berges, N. Tetradis, and C. Wetterich, 
{\it Non-perturbative renormalization flow in quantum 
field theory and statistical physics}, 
Phys. Rep. {\bf 363}, 223 (2002).

\bibitem{diener08} R. B. Diener, R. Sensarma, 
and M. Randeria, 
{\it Quantum fluctuations in the superfluid state of the BCS-BEC crossover}, 
Phys. Rev. A {\bf 77}, 023626 (2008).

\bibitem{obert13} B. Obert, C. Husemann, and 
W. Metzner, 
{\it Low-energy singularities in the ground state of 
fermionic superfluids}, 
Phys. Rev. B {\bf 88}, 144508 (2013).

\bibitem{gorkov61}
L. P. Gorkov and T. K. Melik-Barkhudarov,
{\it Contribution to the theory of superfluidity in an 
imperfect Fermi gas},
Sov. Phys. JETP {\bf 13}, 1018 (1961).

\bibitem{gersdorff01}
G. v. Gersdorff, and C. Wetterich,
{\it Nonperturbative renormalization flow and essential scaling 
for the Kosterlitz-Thouless transition}, 
Phys. Rev. B {\bf 64}, 054513 (2001).

\bibitem{bkt_footnote}
At finite temperatures the proposed truncation is not sufficient to describe the 
relevant BKT physics. Nevertheless, it is clear that also for $T>0$ the flow at large $\Lambda$ is dominated by 
the quantum terms and the initial evolution of the potential proceeds exactly along the same pattern. The additional thermal 
contributions are proportional to $T$ and become relevant exclusively at large RG time (low $\Lambda$) where the structures of
 $U(\alpha)$ typical to a first-order transition had been smeared out by the flow. It is therefore a plausible scenario 
 that a finite BKT transition line (drawn in Fig.~\ref{fig:domes}) will connect the two newly found QCPs.


\bibitem{yamase11}
H. Yamase, P. Jakubczyk, and W. Metzner,
{\it Nematic Quantum Criticality Without Order},
Phys. Rev. B {\b 83}, 125121 (2011).

\bibitem{delmaestro09}
A. Del Maestro, B. Rosenow, and S. Sachdev,
{\it Theory of pair-breaking superconductor-metal 
transition in nanowires}, 
Annals of Physics {\bf 324}, 523 (2009).


\bibitem{bares93}
P.-A. Bares and X.-G. Wen,
{\it Breakdown of the Fermi liquid due to long-range 
interactions}, Phys. Rev. B {\bf 48}, 8636 (1993).

\bibitem{koehl12}
E. Vogt {\it et al.}, 
{\it Scale-Invariance and Viscosity of a Two-Dimensional 
Fermi Gas}, 
Phys. Rev. Lett. {\bf 108}, 070404 (2012).

\bibitem{haussmann09}
R. Haussmann, M. Punk, W. Zwerger,
{\it Spectral functions and rf Response of Ultracold Fermionic Atoms},
Phys. Rev. A {\bf 80}, 063612 (2009).

\bibitem{koehl11a}
M. Feld {\it et al.}, 
{\it Observation of a pairing pseudogap in a 
two-dimensional Fermi gas},
Nature {\bf 480}, 75 (2011).

\bibitem{koehl11b}
B. Froehlich {\it et al.}, 
{\it Radio-Frequency Spectroscopy of a Strongly 
Interacting Two-Dimensional Fermi Gas},
Phys. Rev. Lett. {\bf 106}, 105301 (2011).




\end{thebibliography}
\end{document}